\documentclass[%
 reprint,
nofootinbib,
 amsmath,amssymb,
 aps,
]{revtex4-1}

\usepackage[utf8]{inputenc}

\usepackage{graphicx}
\usepackage{dcolumn}
\usepackage{bm}
\usepackage{hyperref}
\DeclareUnicodeCharacter{2212}{-}

\usepackage{csquotes} 

\usepackage{caption, subcaption}

\usepackage{xcolor}

\usepackage{amsmath}
\usepackage{mathtools}

\usepackage{graphicx}
\usepackage{dcolumn}
\usepackage{bm}
\usepackage{hyperref}

\usepackage{csquotes} 

\usepackage{xcolor}

\renewcommand{\eprint}[1]{\href{http://arxiv.org/abs/#1}{{\tt [arXiv:#1]}}}

\providecommand\jcap{JCAP}

\providecommand\prd{Physical Review D}

\providecommand\mnras{M.N.R.A.S.}

\newcommand{\ie}{{\em i.e.} \/}

\usepackage{graphicx}
\usepackage{caption}
\usepackage{float}

\begin{document}

\title{Constraining decaying dark matter models with gravitational lensing and cosmic voids}


\author{Earl Lester}
\email{earl.sullivanlester@utas.edu.au}
\author{Krzysztof Bolejko}
\email{krzysztof.bolejko@utas.edu.au}
\affiliation{School of Natural Sciences, College of Sciences and Engineering, University of Tasmania, Private Bag 37, Hobart TAS 7001, Australia}

\pacs{98.80.-k, 98.80.Es, 98.80.Jk}


\begin{abstract}
Despite overwhelming observational evidence for dark matter, we still have no evidence of direct detection. Consequently, our knowledge about dark matter is limited, for example, we do not know if dark matter is a stable particle or if it decays. Without a theoretical particle model, the parameter space of possible decay models is highly variable, and astrophysical or cosmological means of indirectly constraining the phenomenological models are required. This paper investigates a scenario in which a dark matter decays and the dark daughter particle moves with respect to the comoving mother particle. The model is parameterised by the decay rate and the injection velocity of the dark matter particles, which can be converted to the mass ratio. In previous work, a simpler model was used to investigate the evolution of cosmic voids typified as regions with low content of galaxies and non-baryonic matter. It was found that the growth of S-type voids is modified by the dark matter decay, leading to imprints at the present day. Here we extend our study and improve the method used to model the  decay. We also study the gravitational weak-lensing signal that will be able to detect or constrain the parameter space of decaying dark matter. The results of this study suggest that future weak lensing surveys may provide unique probes of the phenomological parameters of dark matter.
\end{abstract}

\maketitle

\section{Introduction}

Despite the numerous strengths of the $\Lambda$CDM concordance model of cosmology, the model is not complete with different possible avenues of extensions \cite{bull2016beyond, di2021realm}. Contemporary research suggests that resolutions of some of these discrepancies may be explained by extensions of the standard cold dark matter (CDM) paradigm, particularly through unstable dark matter, \cite{cen2001decaying, pandey2020alleviating}. Recent investigations have appealed to decaying dark matter (DDM) in an effort to resolve present tensions in the Hubble parameter, $H_{0}$, and clustering amplitude, $\sigma_{8}$, observations, \cite{pandey2020alleviating, vattis2019dark}, through possible single-body, \cite{enqvist2015decaying}, or many-body decays, \cite{aoyama2011formulation}.

Assuming that DDM has a lifetime on the order of the age of the universe or greater, then initially there is a correspondence between DDM and the standard CDM. This maintains the timely seeding of structure which motivates CDM but induces modifications in the late-time non-linear growth, particularly to the small-scale power spectrum of matter, similar to  warm dark matter (WDM), \cite{doroshkevich1985fluctuations, wang2014cosmological, boehm2002interacting}. DDM has also been postulated as necessary for the Conformal Cyclic Cosmology, \cite{CCC-DDM}, although in practice the suggested lifetime scale on the order of $10^{11}$ Gy would make it indistinguishable from CDM. 


There exist two predominant types of phenomenological models of invisible decaying dark matter within the literature. In one case, the entirety, or a fraction of the dark matter of the universe is unstable, producing a massless dark radiation via one-body decay, with long lifetimes, \cite{bucko2022constraining}. Some work has shown that these models are inadequate in addressing the present tensions in the concordance model of cosmology, \cite{simon2022constraining}.

In the alternate case, the invisible decay produces a massive daughter particle and a massless radiation via a two-body decay. These models may be parameterised by the decay rate, or lifetime, and mass splitting fraction, or injection velocity. Due to the energy injection these models may be viewed as producing warm dark matter in the late universe, and hence suppressing structure at low redshift, \cite{wang2013lyman}. Due to this property, these versatile models have been proposed to resolve the matter amplitude tension, and additional sub-galactic short-comings of the $\Lambda$CDM models. 

Particle lifetimes in these models range from the age of the universe up to a hundred times this value or more,  \cite{fuss2022decaying}. Furthermore, recent studies have shown evidence for the existence of dark matter decays on much shorter time scales, \cite{Holm_2022}.

This paper tests a middle ground in which the decay lifetime of order of age of the universe $H_{0}^{-1}$ (namely $t_{1/2} = 0.5 H_0^{-1}$, $ t_{1/2} = H_0^{-1}$ and $t_{1/2} = 2 H_0^{-1}$) and explores the possible mass splitting parameter, which is equivalent to the injection velocity imparted to the daughter particle at decay. For the purposes of the study presented here, the effect of the radiative component is assumed to be negligible. 



Due to the difficulty in distinguishing the unique, model-dependent effects of dark matter physics from other standard baryonic processes \cite{2016PhRvD..94l3008L,2018JCAP...07..060C}, in this paper, we aim to investigate the effect of DDM in astrophysical areas with minimal baryonic contamination, \ie cosmic voids, which are large under-dense regions avoided by galaxies. Depending on the choice of tracers, a typical radius of a cosmic void is  $ r_{v} > 15 \, \text{h}^{-1} \text{Mpc}$ \cite{van2011cosmic}. These regions are considered to be the largest structures in the universe and dominate the universal volume, \cite{thompson2011historical}. Assuming a strong mass-to-light ratio, and a correlation between the galaxy morphology and environmental matter density, cosmic voids are primarily composed of DM \cite{peebles2001void}. Therefore, these regions not only minimise baryonic contamination but have less significant dark matter annihilation or other self-interaction effects with quadratic mass dependencies. 

The growth of voids evacuates matter towards the boundary regions, and so the cosmic void under-densities are accompanied by over-dense contiguous shells with a morphology dependent on the void growth and environment, \cite{sheth2004hierarchy}. Consequently cosmic voids are classified as either R-type or S-type, \cite{ceccarelli2013clues}. In the former, the void is in the process of expanding, is embedded in another larger void and has a density profile that rises (R) smoothly to the background density. In contrast, the latter, S-type voids are often smaller surrounded by a compensating shell (S). Typically, small S-type voids tend to be in the process of collapse while the larger R-type are expanding, \cite{sutter2014life}. As it has been shown that an isolated under-density of amorphous shape tend towards spherical form, \cite{icke1984voids}, isolated voids are often modelled as spherical symmetric structures. However, this neglects the complex interactivity of voids between each another and the environment, and the effect this has on their evolution, \cite{sutter2014life}. 

In a previous article, \cite{lester2021imprints}, we investigated the effect of DDM on cosmic void evolution. This work approximated voids as isolated, spherically symmetric under-densities, and showed that the decay of dark matter leads to distinct features in the density profile near the edges of cosmic voids. This showed that by investigating the macro-physical properties of voids, it is possible to constrain the microphysics of dark matter particles.

This paper extends our previous analysis by investigating whether DDM could be constrained by measuring gravitational lensing produced by cosmic voids with DDM. We show that the gravitational lensing signals produced by cosmic voids, \cite{amendola1999weak,krause2012weight}, can be used to probe properties of voids near their edges and thus by doing so, gain better insight into the non-minimal modified dark sector \cite{buckley2018gravitational}.

With the recent first-light of the Rubin telescope and further weak-lensing surveys anticipated to produce unprecedented data,\cite{ivezic2019lsst}, it is a prudent time to explore these avenues of probing non-minimal dark matter models. In this work, we show how our novel model of DDM with massive decay products may be used in conjunction with observations of weak-lensing to explore the phenomenological properties of dark matter. 

The structure of this paper is as follows: Sec. \ref{sec:methods} presents the mathematical formulation of a covariant model of the decay of one dark dust fluid component into another which is taken to model DDM; Sec. \ref{sec:results} presents the results of this model applied to cosmic voids; Sec. \ref{sec:conc} concludes this investigation of the effect of DDM upon the weak-lensing of cosmic voids. 

\section{Methodology} \label{sec:methods}
   
\subsection{The covariant, locally rotationally symmetric framework}

Let us model DDM as two fluids. Each fluid, in its own frame of reference is assumed to be pressure-less dust:
\begin{equation}
    T_{(1)}^{a b} = \rho \, u^{a} u^{b} \quad \text{and} \quad  T_{(2)}^{a b} = \lambda \, v^{a} v^{b} \, \text{.} \nonumber 
\end{equation}

These fluid components are assumed to permeate one another such that interfaces do not form, nor the associated interfacial instabilities \cite{lester2020unstable}. 

The total energy-momentum tensor is the sum of these components. By a Lorentz boost of the form $v^{a} = \gamma (u^{a} + V^{a})$ with a Lorentz factor $\gamma = (1 - V_{a} V^{a})^{-1/2}$ and space-like peculiar velocity $V^{a}$, \cite{king1973tilted}, the total energy-momentum tensor is the $u^{a}$-frame can be re-cast into the usual form for an imperfect fluid. 

Taking the peculiar velocity,  $V^{a}$, to be non-relativistic 
such that  $\gamma \approx 1$, then an observer moving with $u^a$, \ie comoving with the first fluid with density $\rho$, measures the total density to be
\[\mu = \rho + \eta \, \text{,} \]
with $\eta = \gamma^2 \lambda \approx \lambda$, the heat (or energy) flux as
\[ q^{a}  = \eta \, V^{a} \, \text{,} \] and the isotropic and anisotropic pressure terms, $p$ and $\pi^{a b}$, as negligible and may be eliminated. 

Note that for the purposes of this section, we have substituted $\rho$ and $\mu$ for $\kappa c^{2} \rho$ and $\kappa c^{2} \mu$ where $\kappa = 8 \pi G$ and $G$ is Newton's gravitational constant, which is tantamount to adopting geometric units for which $8 \pi G = 1$, and $c = 1$. We re-introduce standard units at the simulation and presentation stages.

The covariant gradient of the 4-velocity field of the first fluid is decomposed into vorticity, $\omega_{a b}$, shear, $\sigma_{a b}$, expansion, $\Theta$, and acceleration, $A_{a}$, \cite{ellis1973relativistic,ellis1999cosmological,tsagas2008}, such that,
\begin{equation}
  \nabla_{b} u_{a} = \omega_{a b} + \sigma_{a b} + \dfrac{1}{3} \Theta \, h_{a b} - A_{a} u_{b}  \, \text{,} \nonumber
\end{equation}
where $h^{a b} = g^{a b} + u^{a} u^{b}$ defines a projection into the instantaneous rest space of an observer.

If the projected spaces are partially iostropic and locally rotationally symmetric (LRS) then there is a preferred space-like unit vector to which all space-like covariant vector fields in the LRS space-time are proportional. Let this preferred space-like unit vector be denoted $z^a$ and note that $z^{a}z_{a} = 1$, and $ u^{a} z_{a} = 0$. The evolution of $z^{a}$ is described by the magnitude of the divergence \[\alpha = D_{a} z^{a} \, \text{,}\] where $D^{a} X = h^{a b} X_{; b}$ is the projected covariant derivative. 


\subsection{Interacting fluids and decaying dark matter}

The interaction between the fluids is modelled as the component-wise violation of conservation of energy-momentum:
\begin{equation}
    \nabla_{b} T^{a b}_{(i)} = I^{a}_{(i)} \, \text{.}
    \label{eq:conservation}
\end{equation}

For this model, the interaction term in Eq. (\ref{eq:conservation}) is given the form
\begin{equation}
I_{(1)}^{a} = - \Gamma \rho (u^a + w^{a}) \, \text{,} \label{eq:interaction}
\end{equation}
where 
\begin{eqnarray}
    w^{a} = \frac{4}{\pi^{2}} v_i \delta z^{a} \, \text{,}
\end{eqnarray}
in which $v_i$ is a scalar that determines the initial injection or kick velocity of the daughter fluid immediately after decay. 

This form of the interaction term, Eq. (\ref{eq:interaction}), is adopted so that the temporal component mimics the standard Rutherford nuclear-type decay, while the spatial component introduces a differential density-dependency. Thus, the parameter $\Gamma$ has an interpretation of the decay rate. With the decay rate, or alternatively the half-life time of order of the age of the universe we have $\Gamma \sim H_{0}$ (as $t_H \sim H_0^{-1}$).

The injection velocity can be represented in terms of the mass splitting parameter
\[ \epsilon = \frac{1}{2} \left( 1 - \frac{m^2}{M^2} \right). \]
Assuming that the decay is into a non-relativistic massive particle and radiation whose energy density is negligible compared to DDM, we have
\[ \epsilon = \frac{1}{2} v_i^2. \]

By the conservation of momentum in the comoving frame, the form of interaction in Eq. (\ref{eq:interaction}) leads to an effective scalar acceleration
\begin{eqnarray}
A = - \frac{4 }{\pi^{2}} \Gamma  \delta \sqrt{2 \epsilon} \, \text{,} \label{eq:acc}
\end{eqnarray}
through which the two decay parameters become degenerate. Furthermore, note that the transfer of energy-momentum between the two fluids, which models the decay, has led to world-lines that are not time-like geodesics.

\subsection{Evolution equations}

We refer the reader to our previous paper, \cite{lester2021imprints}, for further details of the assumptions and derivations. In that work we showed that they lead to a system of propagation equations:
\begin{align}
&\dot{\Theta} = -\dfrac{1}{3} \Theta^{2} - \dfrac{1}{2} \mu - \frac{3}{2} \Sigma^2 + A \alpha + A^2 + A' + \Lambda	\,\text{,} \label{eq:raych_eq} \\
&\dot{ \Sigma} = - \dfrac{2}{3} \Theta \Sigma - \frac{1}{2} \Sigma^{2} - {\cal W} - \frac{1}{3} \alpha \, A + \frac{2}{3} A^{2} + \dfrac{2}{3} A' \,\text{,} \label{eq:scalar_shear} \\
&\dot{{\cal W}}  =  - \Theta {\cal W} - \dfrac{1}{2} \mu  \Sigma + \frac{3}{2} \Sigma {\cal W} + \frac{1}{6} \alpha \, Q  - \frac{1}{3} Q'  - \frac{2}{3} A Q	\,\text{,} \label{eq:scalar_weyl} \\
&\dot{\mu} = -\Theta \mu  - Q' - \alpha \, Q - 2 A Q	\,\text{,}  \label{eq:scalar_mu} \\
&\dot{\rho} = -\Theta \rho - \Gamma \rho \,\text{,}  \label{eq:scalar_rho} \\
&\dot{Q} = - (\Sigma + \frac{4}{3} \Theta) Q - \mu \, A	\,\text{,} \label{eq:heat_prop}
\end{align}
where $Q, A, \Sigma, {\cal W}$ are the scalar magnitudes of the LRS heat and acceleration vectors, and shear and electric Weyl tensors, respectively, and where $\dot{X} = u^{a} \nabla_{a} X$ and $X' = z^{a} \nabla_{a} X$.

To complete this system of evolution equations, we require
\begin{equation}
    \Theta = \frac{\dot{\nu}}{\nu} \quad \text{ and} \quad \alpha + A = \frac{\nu'}{\nu} \,\text{,} \label{eq:nu_eqs}
\end{equation} 
where $\nu$ is the volume element and $\nu = \sqrt{-g}$, with $g$ representing the metric determinant. 

\subsection{Initial conditions}\label{initialLTB}

Initially, it is assumed there exists only a single, comoving dust fluid, so that $\eta = 0$ and $V = 0$ and the system is approximated by an exact spherically symmetric solution single dust, \ie the Lema\^itre--Tolman model.

We then use this model to specify the initial conditions for the evolution equations of $\Theta$, $\Sigma$, ${\cal W}$, $\mu$, $\rho$, and $Q$. The Lema\^itre--Tolman model is in turn specified by two restrictions: (i) the age of the universe is uniform, $t_B = 0$, and (ii) a mass profile of form:
\begin{equation}
M(r) = \frac{1}{6}  \rho_{i} \left[ 1 + \frac{1}{2} m_0 \left( 1 - \tanh \frac{r-r_0}{2 \Delta r} \right)  \right] r^3, \label{eq:ltb_m_ic}
\end{equation} 
with
\begin{equation}
\rho_{i} = \frac{ 3 H_0^2 \Omega_M}{8 \pi} (1+z_{i})^3 . \label{eq:bg_density}
\end{equation}

The choice of the hyperbolic tangent profile in Eq. (\ref{eq:ltb_m_ic}) allows one to model a single under-density and its asymptotical behaviour, and ensures that for large $r$ the model approaches the FLRW cosmological background. 

The initial redshift was chosen to be $z_{i} = 150$ as this ensures that the ratio of energy densities of radiation and matter is negligible, that is
\begin{eqnarray}
    \frac{\epsilon_{R}}{\epsilon_{M}} = \frac{\Omega_{R}}{\Omega_{M}}(1 + z) < 0.01 \nonumber \, \text{,}
\end{eqnarray}
which is required by the assumed non-relativistic speeds of the fluid components in this model. Note that in a flat $\Lambda$CDM model, this redshift corresponds to an age of approximately $8.9 \, \text{Myr}$ and hence prior to this time no significant decay has occurred, assuming decay rates of the order of $H_{0}$. 

\begin{figure*}
    \begin{center}
      \includegraphics[scale=0.72]{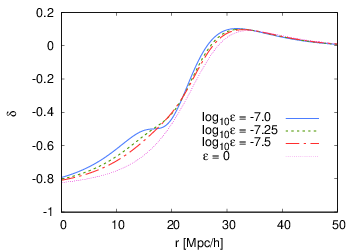}
    \end{center}
    \caption{Density contrast of a void located at $z=0.25$ with parameters $m_0 = -0.034$, $r_0=130$ Kpc, and $\Delta r=71.5$ Kpc and injection velocity $v_i = 0 \, \text{km} \text{s}^{-1}$ ($\epsilon = 0$), $v_i = 75$ km/s ($\log_{10} \epsilon = -7.5$), $v_i = 100\, \text{km} \text{s}^{-1}$($\log_{10} \epsilon = -7.25$), and $v_i = 135 \, \text{km} \text{s}^{-1}$ ($\log_{10} \epsilon = -7.0$).}
    \label{fig1}
\end{figure*}

The initial profile of Eq. (\ref{eq:ltb_m_ic}) includes three free parameters: $m_0$, $\Delta r$, and $r_0$. These parameters were randomised for a Monte-Carlo simulation and set to be
\[ m_0 = -0.0225 + 0.0075 \cdot {\cal U}_{[0,1]},\]
where ${\cal U}_{[0,1]}$ is a number drawn from a uniform distribution between 0 and 1, and
\[ r_0 = \left( 45 + 155 \cdot {\cal U}_{[0,1]} \right) [{\rm Kpc}],    \]
and
\[\Delta r = \left( 0.425 + 0.075 \cdot {\cal U}_{[0,1]} \right) \cdot r_0.  \]
The reason for this particular choice is justified in Sec. \ref{fine-tuning}. On prescribing the initial conditions it is possible to evolve the scalar system (\ref{eq:raych_eq}) - (\ref{eq:nu_eqs}) from the given initial redshift to any desired later redshift representing the position of the void as a lens. An example of a present-day profile of a void parameterised 
with $m_0 = -0.034$, $r_0=130$ Kpc, and $\Delta r=71.5$ Kpc is presented in Fig. \ref{fig1}.

\subsection{The void size function}

\begin{figure*}
    \begin{center}
      \includegraphics[scale=0.55]{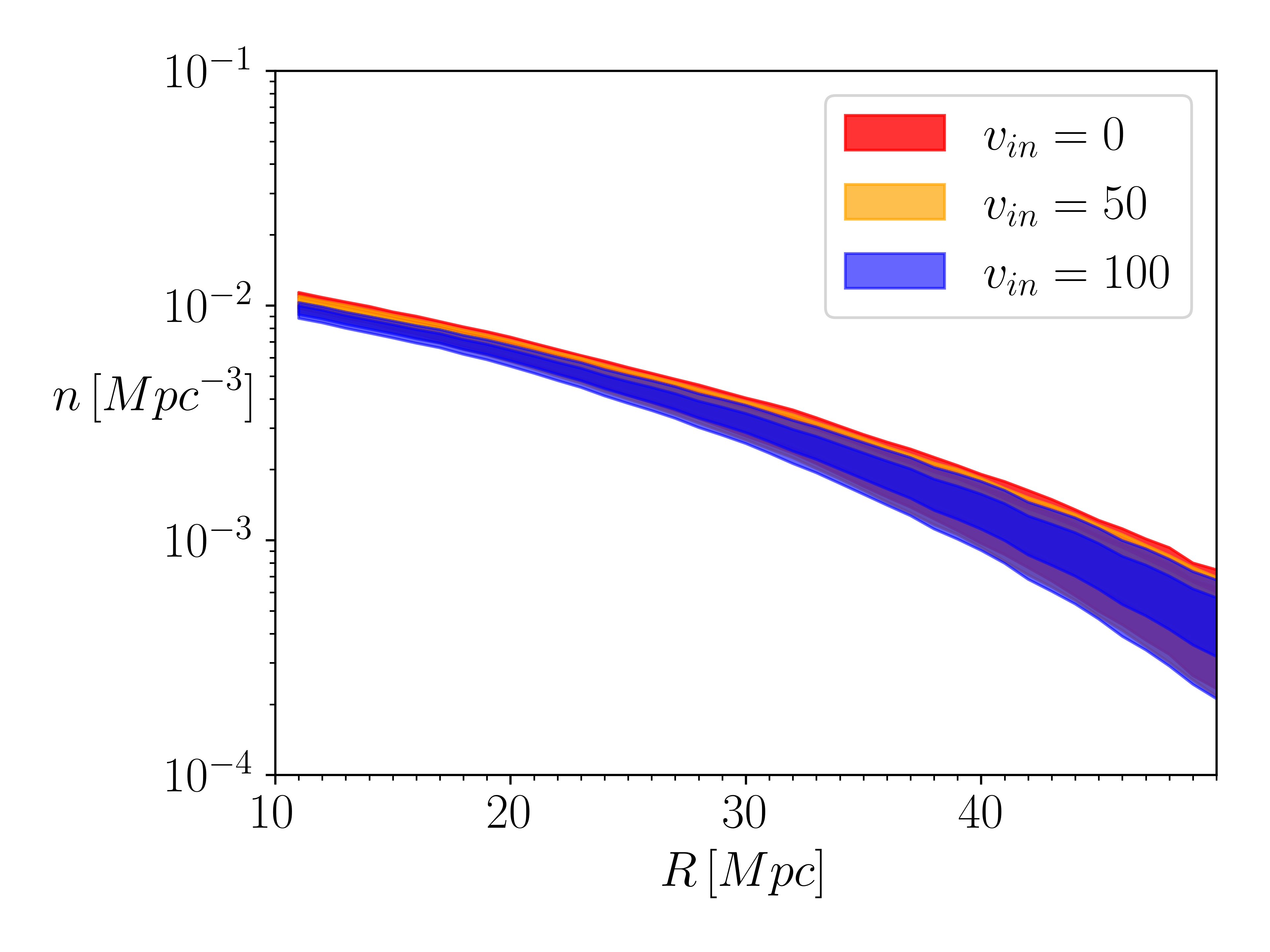}
    \end{center}
    \caption{The void size function with the 68 \% and 95 \% bands for three cases of increasing injection velocity, $v_{in} = 0$, $50$ and $100 \text{km}\text{s}^{−1}$, with the decay rate set to $\Gamma = \text{H}_{0}$. The results show that there are clearly no significantly non-overlapping regions, and therefore that a strong signature of dark matter decays cannot be determined by this approach.}
    \label{fig_vsf}
\end{figure*}

The void size function (VSF) has been determined for recent surveys, and is anticipated to be an important cosmological probe in future surveys, \cite{contarini2022euclid, contarini2023cosmological}. In Figure \ref{fig_vsf} the effect of dark matter decays upon the void size function is demonstrated. This figure shows the fractional volume of voids of given radii expected in a mock void catalogue in the redshift range of $z = 0.2$ to $z = 0.3$, evolved from $z = 150$, for three cases of the decaying dark matter model developed here with decay parameters of $\Gamma = H_{0}$ and $v_{in} = 0, 50$ and $100 \text{km}\text{s}^{−1}$. 

The voids, defined by as under-densities between $-0.95 \leq \delta \leq -0.5$, are binned by $1 \, \text{Mpc}$ radii. For each sample, the standard deviation of the density field is determined. For each bin, samples are drawn from a normal distribution. The number density, $n$, is calculated as the number of voids in each radius bin. This process is repeated until the volume matches the volume of the survey. It is clear from Figure \ref{fig_vsf}, which shows the number density $n$ against the radii, $R$, that the injection velocity increases do not lead to a statistically significant difference in the VSF.
Therefore the VSF does not seems as a promising tool to detect possible signatures of the dark matter decay. However, as see in Figure. \ref{fig1}, while the size of the void is not affected significantly, its depth and gradient is also affected. We next explore whether the gravitational lensing (which is sensitive to both depth and gradient) could be used as a tool to detect dark matter decay signatures. 

\subsection{Weak gravitational lensing}

Gravitational lensing leads to a discrepancy between the actual position of a source (in the absence of a lens) and the apparent image. In astrophysical situations of interest there is a small deflection angle between the angular position of the source and the observed image. Note that the deflection is negative for under-densities. Because of this small deflection angle, linearised general relativistic theory is a valid approximation, \cite{blandford1992cosmological}.

The Jacobian between the unlensed (i.e. homogeneous case) and apparent (i.e. lensed case) source positions is reduced to a $2 \times 2$ distortion or magnification matrix, \cite{bartelmann2010gravitational}. This assumes that the source and image belong to the same sub-space normal to the photon paths and observer 4-velocity \cite{bonvin2008effect}. When this is not the case Doppler magnification should be accounted for, but the present investigation assumes the simplest case in which these effects can be ignored. The distortion of magnification matrix can be parameterised by the scalar convergence, $\kappa$, and tangential shear, $\gamma$, \cite{bartelmann2001weak}.

The lensing convergence, $\kappa$, corresponds to an isotropic (de-)magnification of the source images, while the shear, $\gamma$, is the trace-free part of the distortion matrix, which describes the tidal gravitational forces, and anisotropically distorts the shape of the image. The determinant of the inverse magnification matrix is the magnification scalar $\mu = ((1 - \kappa)^{2} - |\gamma|^2)^{-1}$. The weak lensing limit is achieved if the lensing convergence and shear $\kappa, \gamma << 1$ and the magnification is then approximately 
\begin{equation}
    \label{eq:lin_mag}
    \mu = 1 +2 \kappa \, \text{.}
\end{equation}

When the lens possesses azimuthal symmetry, the tangential shear is proportional to the differential surface mass density of the lens, while the convergence is proportional to the projected surface mass density of the lens. The projected, or surface, mass density $\Sigma$ is the component projected onto a plane perpendicular to the light-ray and is defined by
\begin{equation}
\Sigma = \int  \bar{\rho} \, \delta(r(\theta,z)) \, d \ell  \, \text{.} \label{eq:sigma_int}
\end{equation}
Here $r(\theta,z)$ is the radius from observer to lens centre, $\theta$ is the angle from perpendicular to lens plane (\ie $\theta = 0$ specifies the void centre), $\bar{\rho}$ is the background density given in Eq. (\ref{eq:bg_density}), $\ell$ is proper distance along the line-of-sight, and $\delta$ is the density perturbation determined by the void density profile. Note that this quantity shares the same name and notaional representation as the previously defined shear of a fluid element, as it is the shear of a congruence of null-geodesics, \cite{sasaki1993cosmological}. The convergence is then
\begin{equation}
     \kappa = \dfrac{\Sigma}{\Sigma_{C}} \, \text{,} \label{eq:kappa_defn}
\end{equation}
and, assuming azimuthal symmetry, the shear is
\begin{equation}
    \gamma = \dfrac{\bar{\Sigma} - \Sigma}{\Sigma_{C}} \, \text{,} \label{eq:gamma_def}
\end{equation}
where $ \bar{\Sigma}$ is the differential surface mass density, and $\Sigma_{C}$ is critical mass density given by
\begin{equation}
    \Sigma_C = \dfrac{1}{4 \pi} \dfrac{D_s}{D_l D_{sl}} \, \text{.} \nonumber
\end{equation} 
Here $D_{i j} = D_A(z_i,z_j)$ is the proper angular distance between two redshifts, which is the ratio the transverse to angular sizes of an object, and $D_{i} = D_A(0,z_j)$, not to be confused with the spatially-projected covariant derivative defined above. The critical surface mass density defines the threshold for the lens to be considered strong, in which case the lens may then possibly produce multiple images, \cite{bartelmann2001weak}.

\begin{figure*}
    \begin{center}
      \includegraphics[scale=0.45]{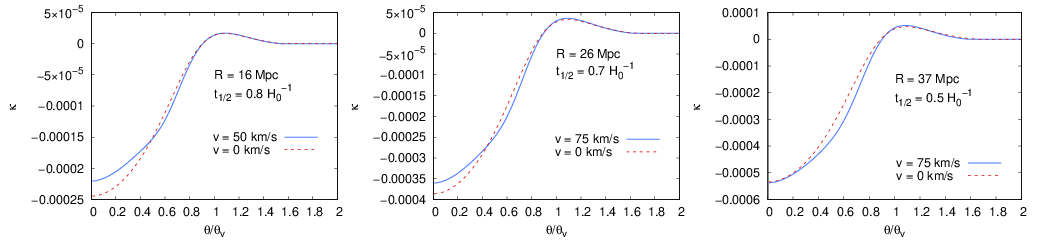}
    \includegraphics[scale=0.45]{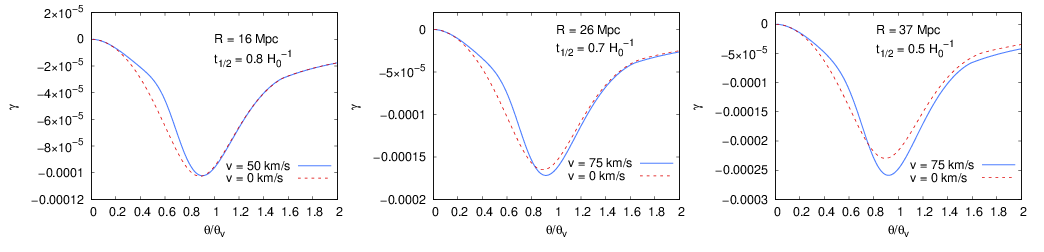}
    \end{center}
    \caption{Weak lensing convergence (upper panels) and shear (lower panels) of a void located at $z=0.25$ with parameters increasing size and decay rate (from left to right).}
    \label{WL}
\end{figure*}

In order to probe the WL signal produced by voids in the presence of DDM we compute the differential and projected surface mass densities by numerical integration of Eq. (\ref{eq:sigma_int}) over the required proper distance for a range of $\theta$. At each $\theta$ the density perturbation is found from the given void profile that is determined by numerical integration of (\ref{eq:raych_eq}) - (\ref{eq:heat_prop}) along with (\ref{eq:nu_eqs}) from given initial conditions specified by (\ref{eq:ltb_m_ic}), at the initial instant that corresponds $z = 150$. A method of lines for evolving the partial differential equations was adopted and the numerical integration of the system of differential equations was performed using the Dormand--Prince scheme. A few examples of a weak lensing signal of voids of different size located at $z_l= 0.25$ and a source at $z_s = 0.5$ is presented in Figs. \ref{WL}. 

\subsection{Universal void profile}\label{HSW-profile}

The universal profile of a cosmic void is the HSW profile which is determined empirically from $N$-body simulations, given by, \cite{hamaus2014universal},
\begin{equation}
\delta (r) = \delta_{c} \frac{1 - (r/r_{s})^{\alpha}}{1 + (r/r_{v})^{\beta}}
\end{equation}
where $\delta_{c}$ is the central density contrast, $\alpha$ and $\beta$ are slope parameters, and $r_{s}$ and $r_{v}$ determine the size of void and shell. The parameters $\alpha$, $\beta$, $\delta_c$ and $r_s$ were specified 
as per Fig. 2 of \cite{hamaus2014universal}. The parameter $\alpha$
\[\alpha = {\cal N} (\mu = \bar{\alpha}, \sigma = 0.15), \]
where ${\cal N} (\mu = \bar{\alpha}, \sigma = 0.15)$ is a 
random number drawn from the normal distribution with the mean
\[ \bar{\alpha} = -2 \frac{r_s}{r_v} + 4.0, \]
and standard deviation $0.15$. The parameter $\beta$ 
\[\beta = {\cal N} (\mu = \bar{\beta}, \sigma = 0.15), \]
\[
\bar{\beta} = \begin{dcases*}
17.5 \frac{r_s}{r_v} - 6.5   & for \, $r_v < 17.5 $ Mpc/h \\[1ex]
-9.8\frac{r_s}{r_v}   + 18.4
   & for \, $r_v \ge 17.5 $ Mpc/h ;
\end{dcases*}
\]
the parameter $\delta_c$ 
\[
\delta_c = \begin{dcases*}
 -0.92 + 0.05 \cdot {\cal U}_{[0,1]}   & for \, $r_v < 17.5 $ Mpc/h \\[1ex]
 -0.88 + 0.05 \cdot {\cal U}_{[0,1]} 
   & for \, $r_v \ge 17.5 $ Mpc/h
\end{dcases*}
\]
and the parameter $r_s$ 
\[
r_s = \begin{dcases*}
\left( 0.87 + 0.03 \cdot {\cal U}_{[0,1]} \right) r_v    & for \, $r_v < 17.5 $ Mpc/h \\[1ex]
\left( 0.89 + 0.03 \cdot {\cal U}_{[0,1]} \right) r_v 
   & for \, $r_v \ge 17.5 $ Mpc/h \, \text{.}
\end{dcases*}
\]

\subsection{Fine-tuning the free parameters}\label{fine-tuning}

The aim of this paper is to investigate the weak gravitational signal of DDM near the edges of cosmic voids. The gravitational lensing signal increases with redshift of the lens and source, however, the DDM features only becomes prominent at low redshifts, \cite{lester2021imprints}. 

In order to maintain a minimal volume that allows us to detect sufficient voids to stack the signal the lens should be at lower redshift than $z \approx 0.2$ \cite{hossen2022mapping,hossen2022ringing}. We thus place the center of the void at a random redshift
\[z_{void} ={\cal N} (\mu = 0.25, \sigma = 0.025),  \]
where ${\cal N} (\mu = 0.25, \sigma = 0.025)$ is a random number drawn from the normal distribution with mean $\mu = 0.25$ and standard deviation $\sigma = 0.025$. Similarly we place the source at a random redshift
\[z_{source} ={\cal N} (\mu = 0.5, \sigma = 0.05). \]

\begin{figure*}
    \begin{center}
      \includegraphics[scale=0.45]{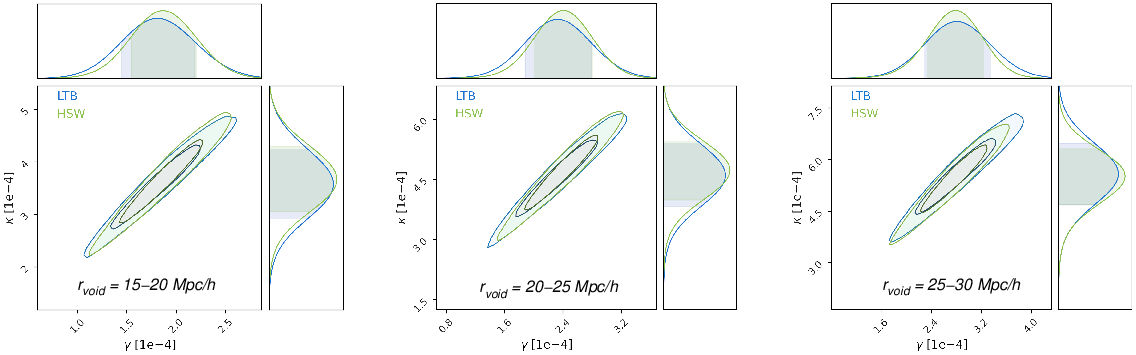}
    \end{center}
    \caption{The distribution of values of maximal amplitude of weak lensing convergence and maximum value of weak lensing shear.
    The weak lensing was evaluated using two types of voids: voids that are given by an LTB solution of Sec. \ref{initialLTB} 
    and by the HSW \ref{HSW-profile}.}
    \label{fig4}
\end{figure*}

From Eq. (\ref{eq:kappa_defn}) and (\ref{eq:gamma_def}), we determine the convergence and shear signal produced by a void which is described using HSW profile of Sec. \ref{HSW-profile} as well as a void in the case of no decay (\ie the Lema\^itre--Tolman model of Sec. \ref{initialLTB}). The results are presented in Fig. \ref{fig4}. These figures present justification for the choice of the initial parameters $m_0$, $\Delta r$, and $r_0$ of Sec.  \ref{initialLTB}. These parameters were fine-tuned to recover the properties of the voids as given by the HSW profile, \cite{hamaus2014universal}. 

\section{Results} \label{sec:results}


\begin{figure*}
\centering
\begin{subfigure}[h]{0.3\textwidth}
    \centering
    \includegraphics[scale=0.45]{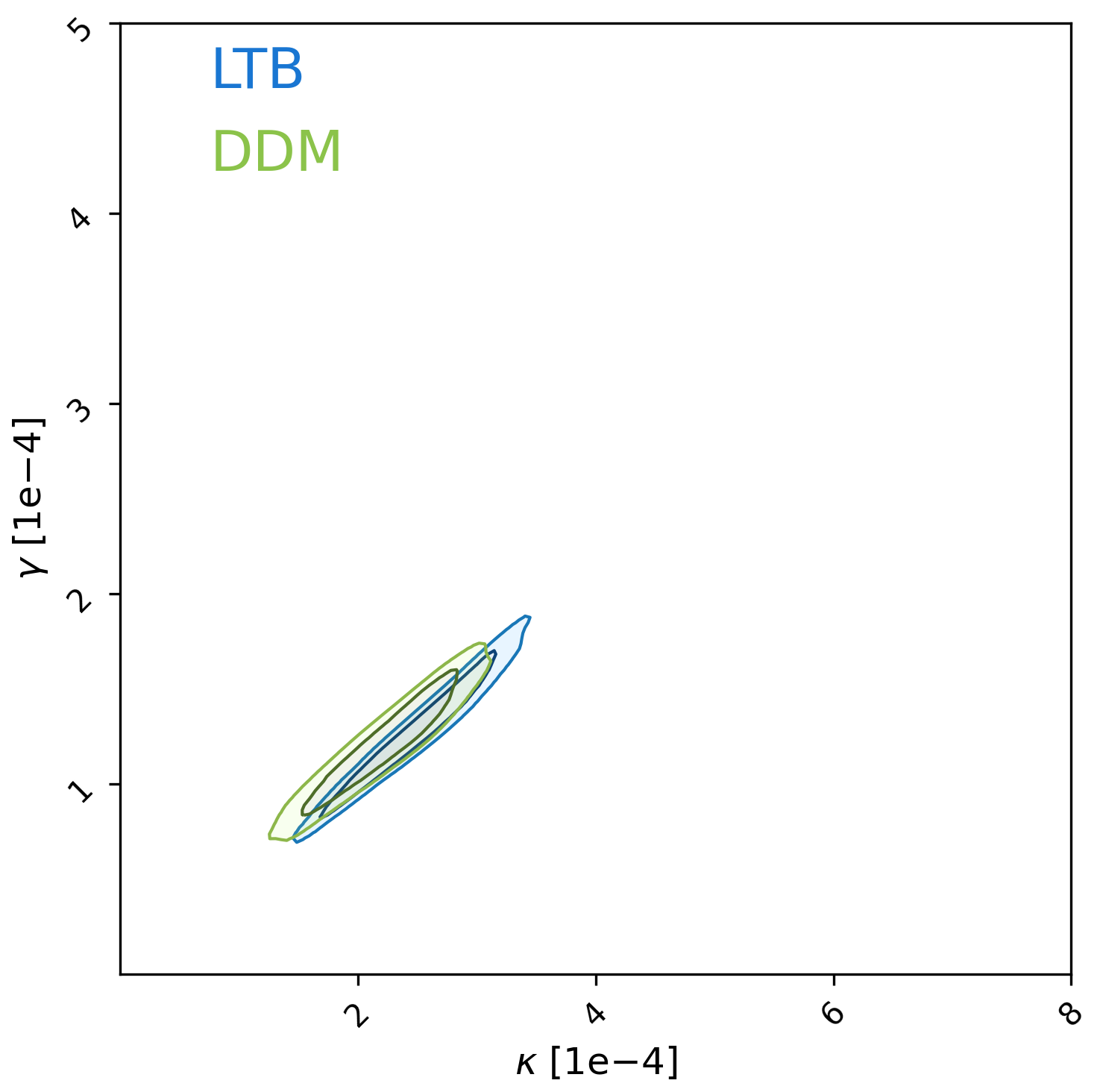}
    \caption{$10 < R_{v} < 20$ Mpc, $v_{in} = 25 \, \text{km}{s}^{-1}$}
    \label{subfig:1.1}
\end{subfigure}%
~ 
\begin{subfigure}[h]{0.3\textwidth}
    \centering
    \includegraphics[scale=0.45]{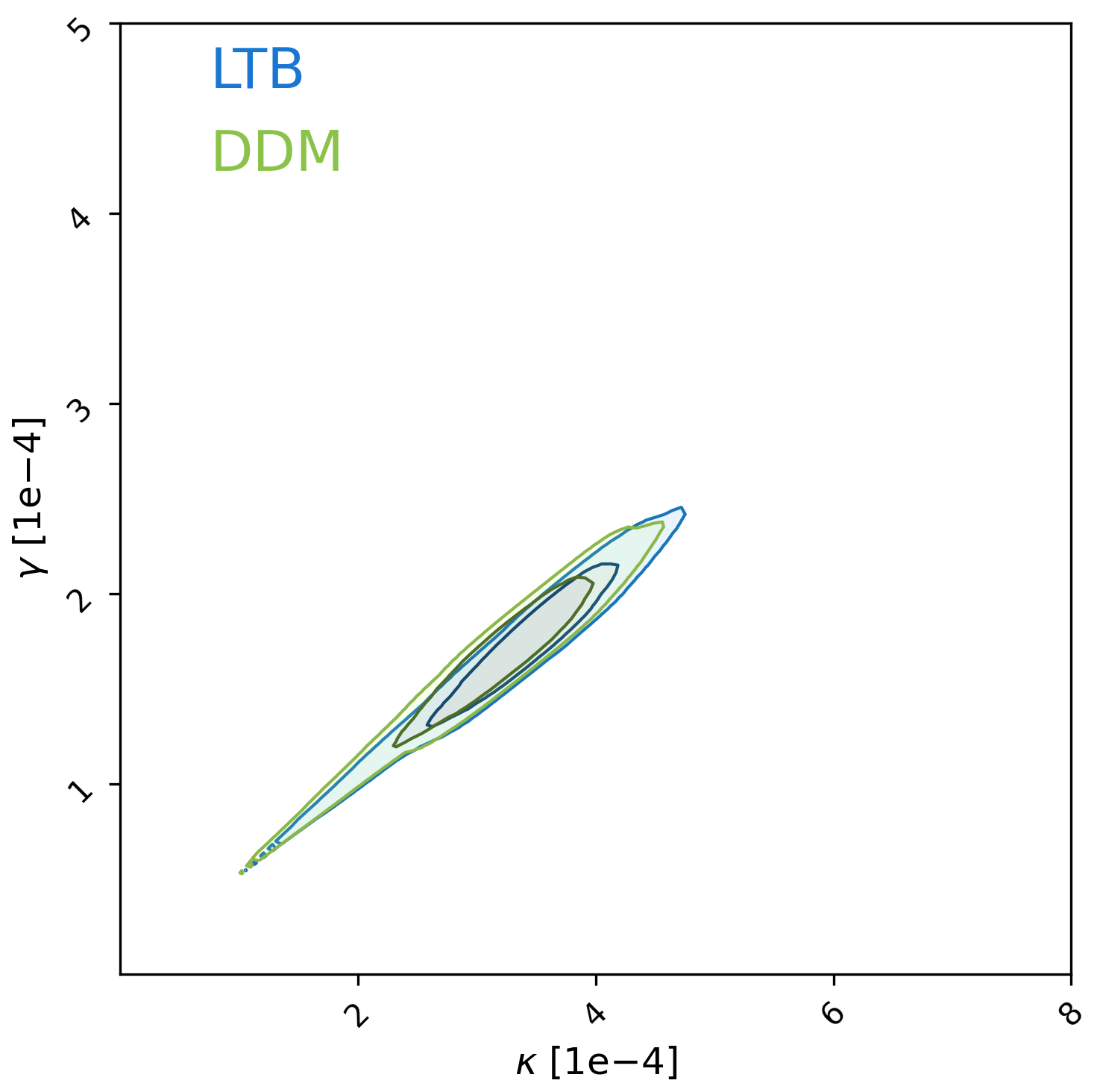}
    \caption{$20 < R_{v} < 30$ Mpc, $v_{in} = 25 \, \text{km}{s}^{-1}$}
    \label{subfig:1.2}
\end{subfigure}
~ 
\begin{subfigure}[h]{0.3\textwidth}
    \centering
    \includegraphics[scale=0.45]{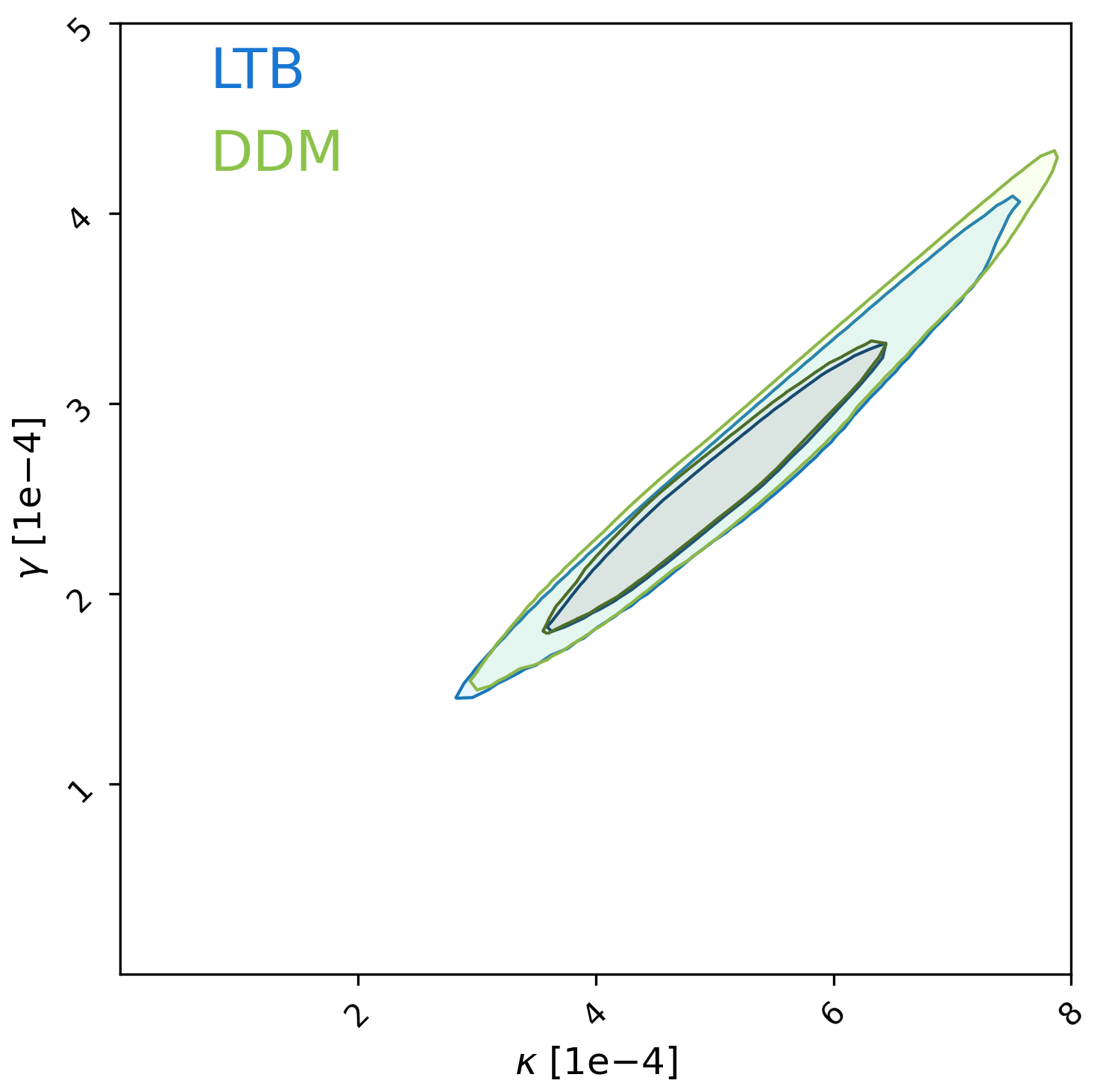}
    \caption{$30 < R_{v} < 40$ Mpc, $v_{in} = 25 \, \text{km}{s}^{-1}$}
    \label{subfig:1.3}
\end{subfigure}
\vfill
\begin{subfigure}[t]{0.3\textwidth}
    \centering
    \includegraphics[scale=0.45]{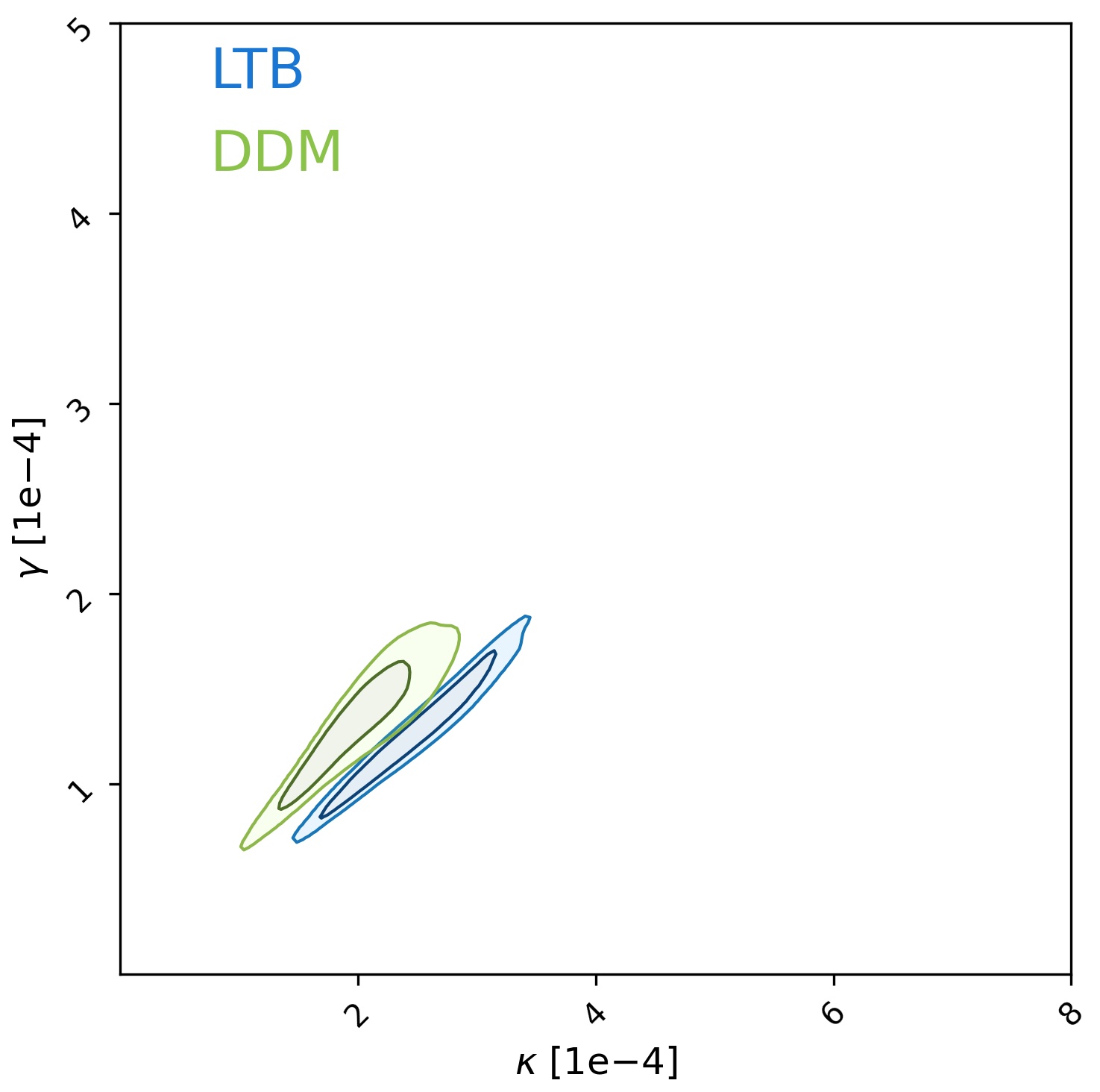}
    \caption{$10 < R_{v} < 20$ Mpc, $v_{in} = 75 \, \text{km}{s}^{-1}$}
    \label{subfig:1.4}
\end{subfigure}%
~ 
\begin{subfigure}[t]{0.3\textwidth}
    \centering
    \includegraphics[scale=0.45]{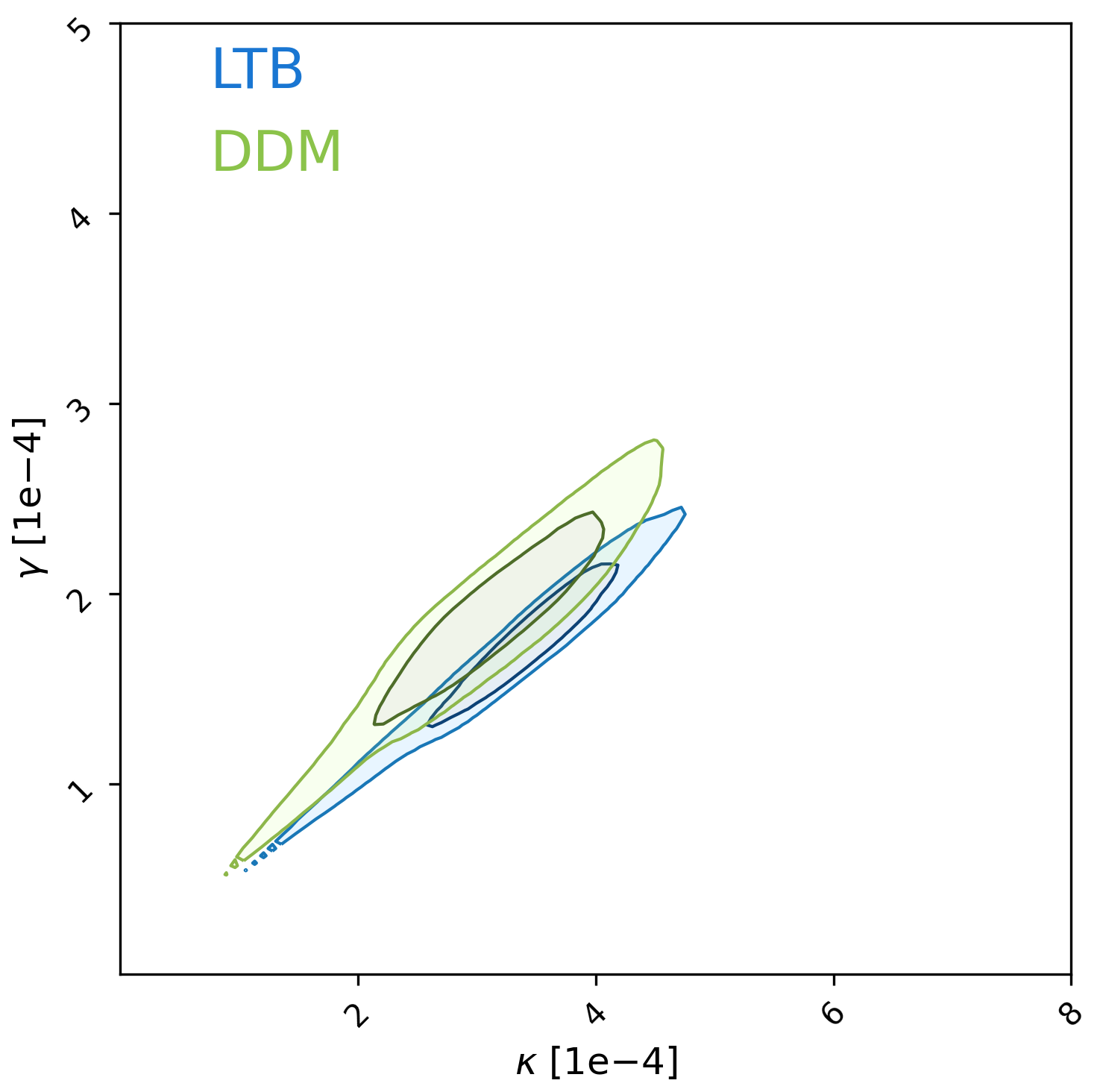}
    \caption{$20 < R_{v} < 30$ Mpc, $v_{in} = 75 \, \text{km}{s}^{-1}$}
    \label{subfig:1.5}
\end{subfigure}
~ 
\begin{subfigure}[t]{0.3\textwidth}
    \centering
    \includegraphics[scale=0.45]{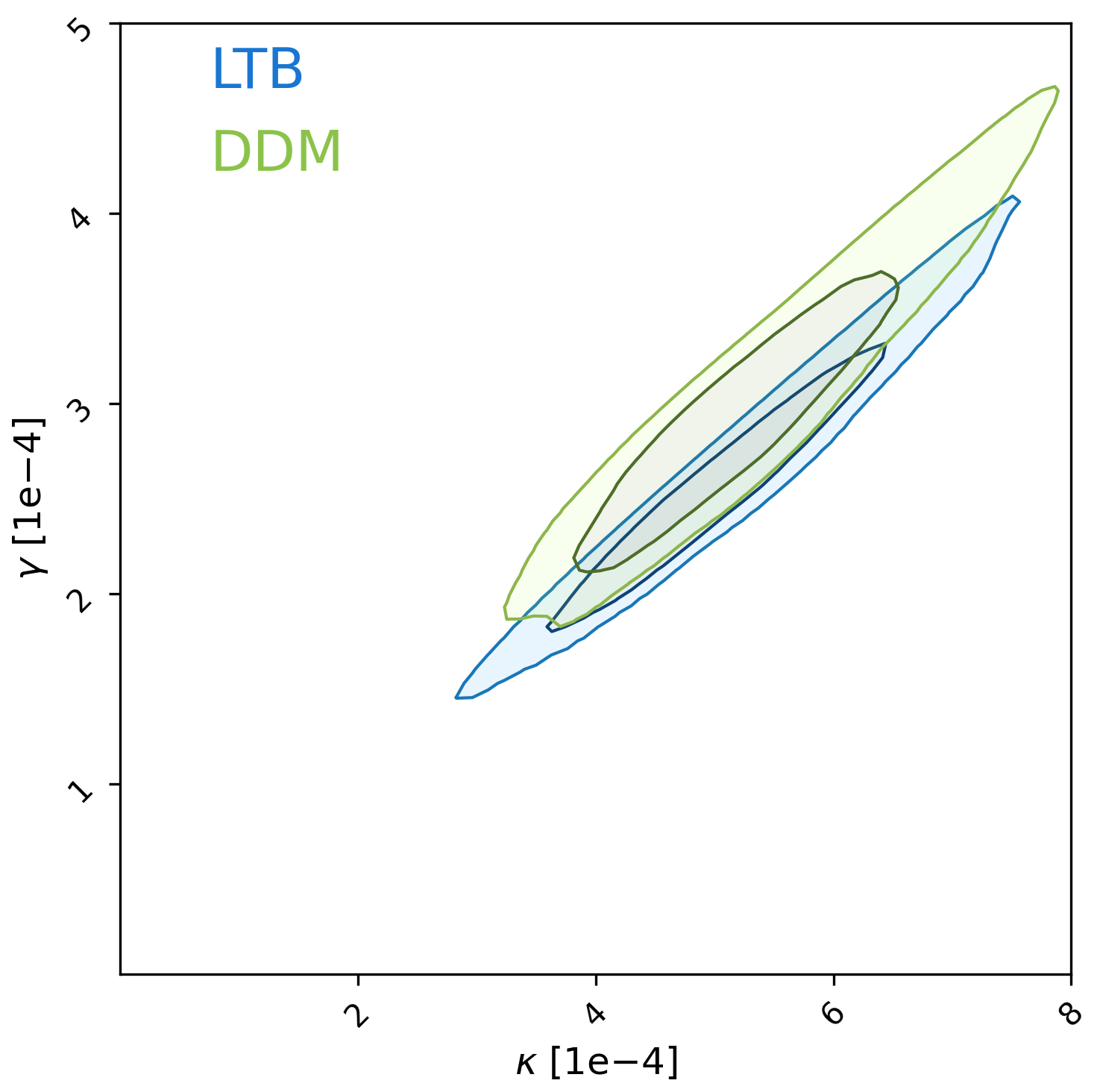}
    \caption{$30 < R_{v} < 40$ Mpc, $v_{in} = 75 \, \text{km}{s}^{-1}$}
    \label{subfig:1.6}
\end{subfigure}
\caption{The distribution of values of maximal amplitude of weak lensing convergence and maximum value of weak lensing shear for voids with DDM and without DDM (i.e. the LTB case), for the three void size classes, the low decay-rate case of $\Gamma = 0.5 \, \text{H}_{0}$ and two injection velocities $v_{in} = 25$ and $75 \, \text{km}\text{s}^{-1}$.}
\label{fig:fig_wl_g05}
\end{figure*}

\begin{figure*}
\centering
\begin{subfigure}[h]{0.3\textwidth}
    \centering
    \includegraphics[scale=0.45]{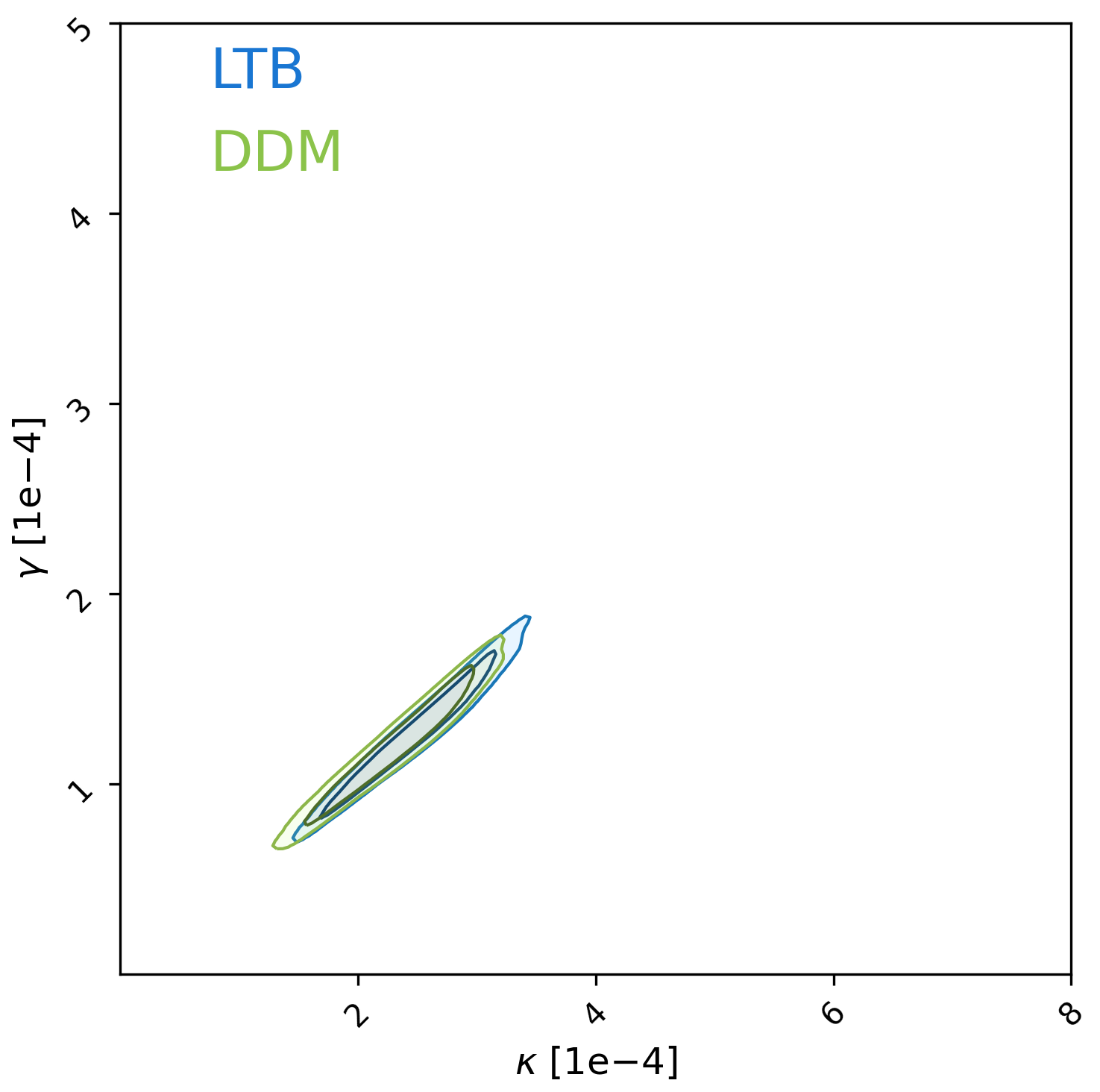}
    \caption{$10 < R_{v} < 20$ Mpc, $v_{in} = 25 \, \text{km}{s}^{-1}$}
    \label{subfig:2.1}
\end{subfigure}%
~ 
\begin{subfigure}[h]{0.3\textwidth}
    \centering
    \includegraphics[scale=0.45]{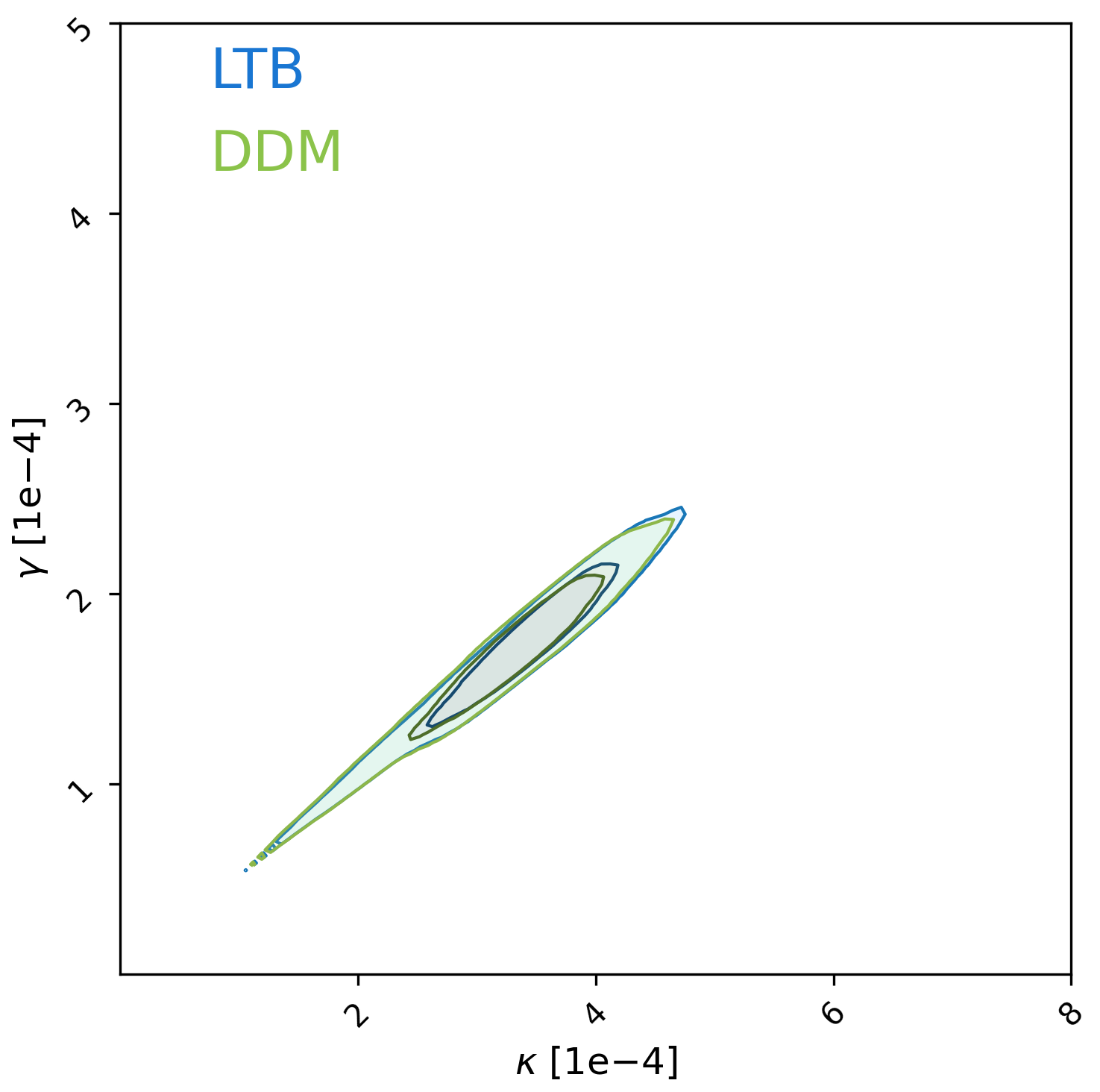}
    \caption{$20 < R_{v} < 30$ Mpc, $v_{in} = 25 \, \text{km}{s}^{-1}$}
    \label{subfig:2.2}
\end{subfigure}
~ 
\begin{subfigure}[h]{0.3\textwidth}
    \centering
    \includegraphics[scale=0.45]{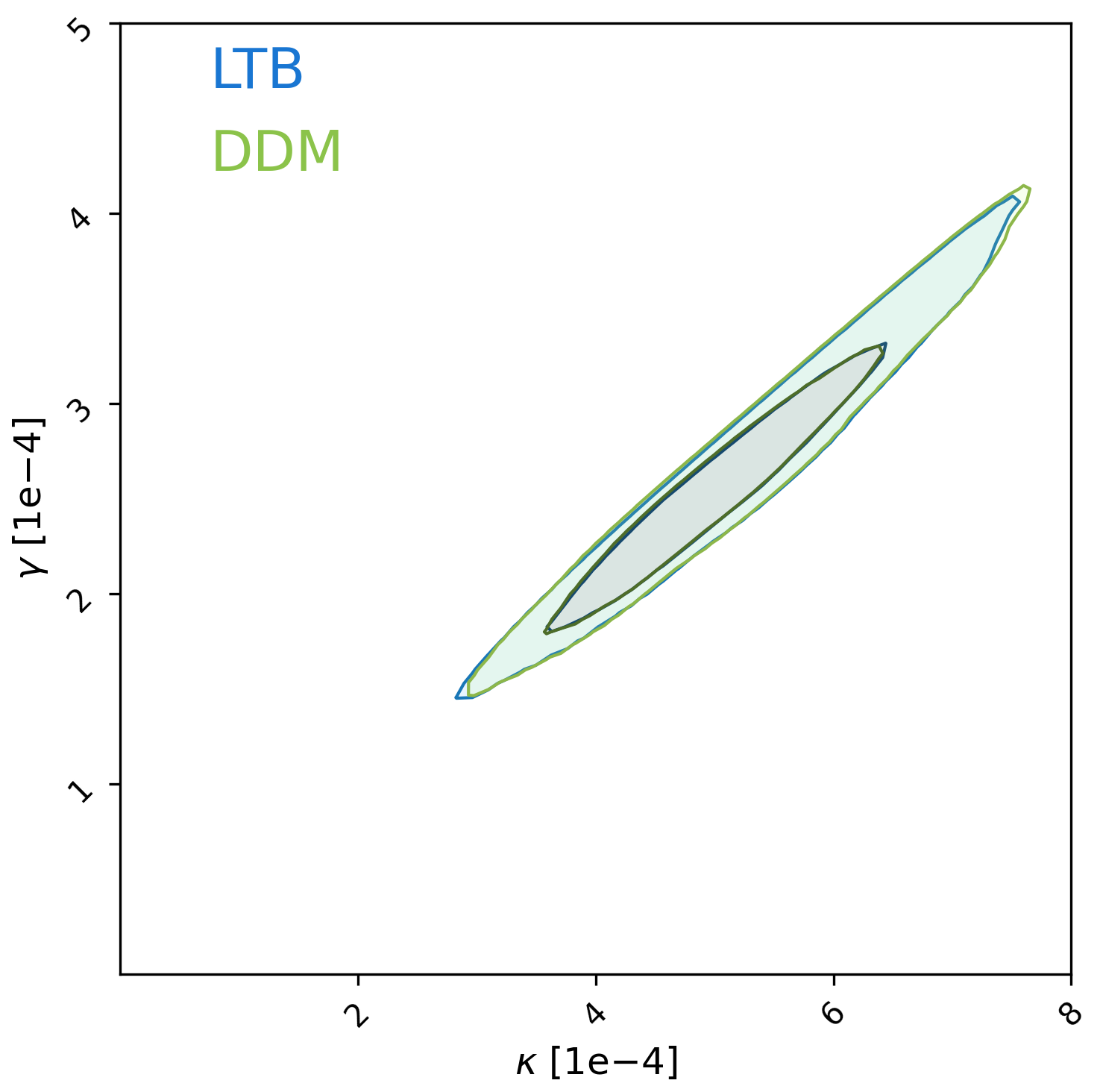}
    \caption{$30 < R_{v} < 40$ Mpc, $v_{in} = 25 \, \text{km}{s}^{-1}$}
    \label{subfig:2.3}
\end{subfigure}
\vfill
\begin{subfigure}[t]{0.3\textwidth}
    \centering
    \includegraphics[scale=0.45]{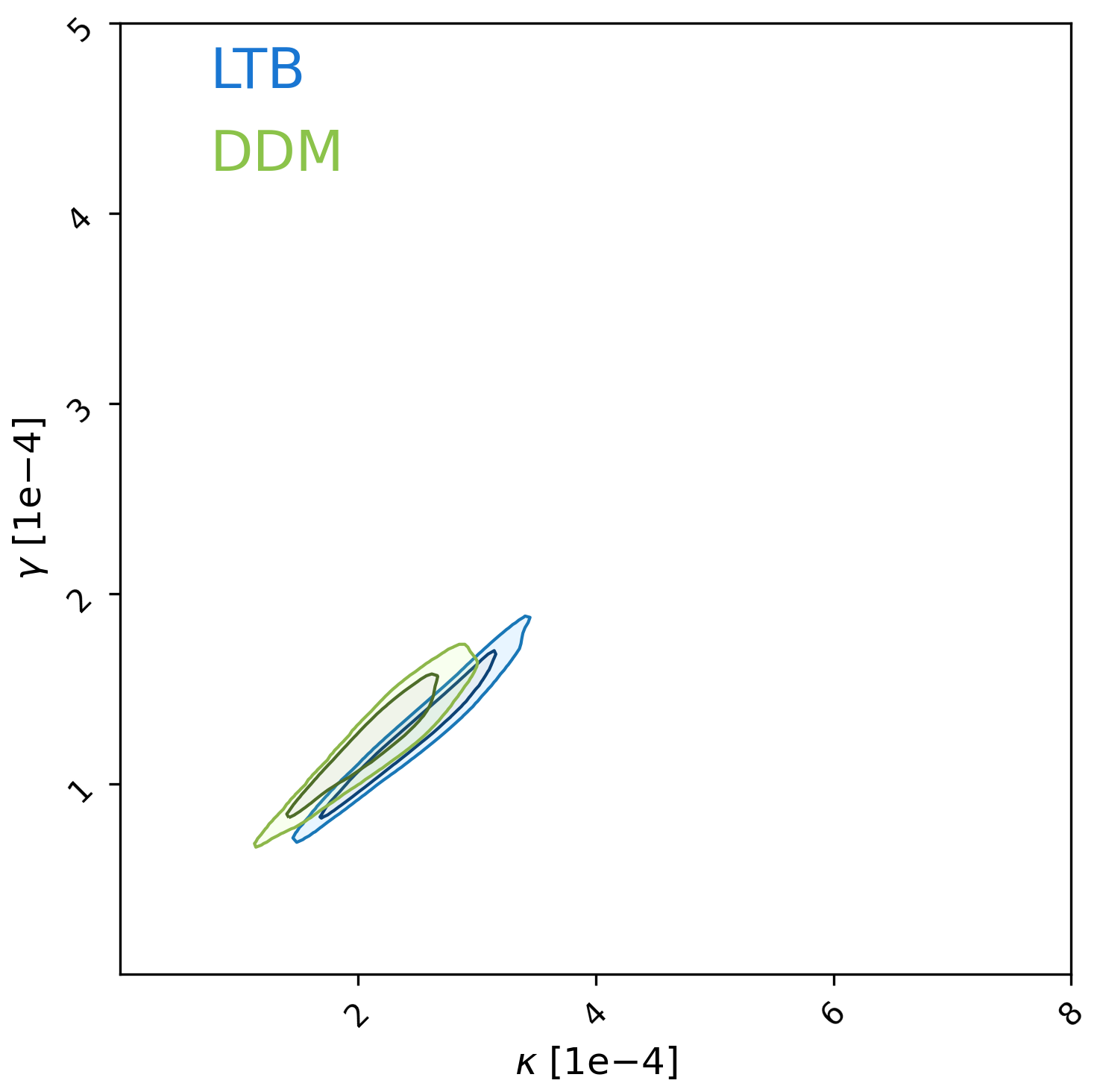}
    \caption{$10 < R_{v} < 20$ Mpc, $v_{in} = 75 \, \text{km}{s}^{-1}$}
    \label{subfig:2.4}
\end{subfigure}%
~ 
\begin{subfigure}[t]{0.3\textwidth}
    \centering
    \includegraphics[scale=0.45]{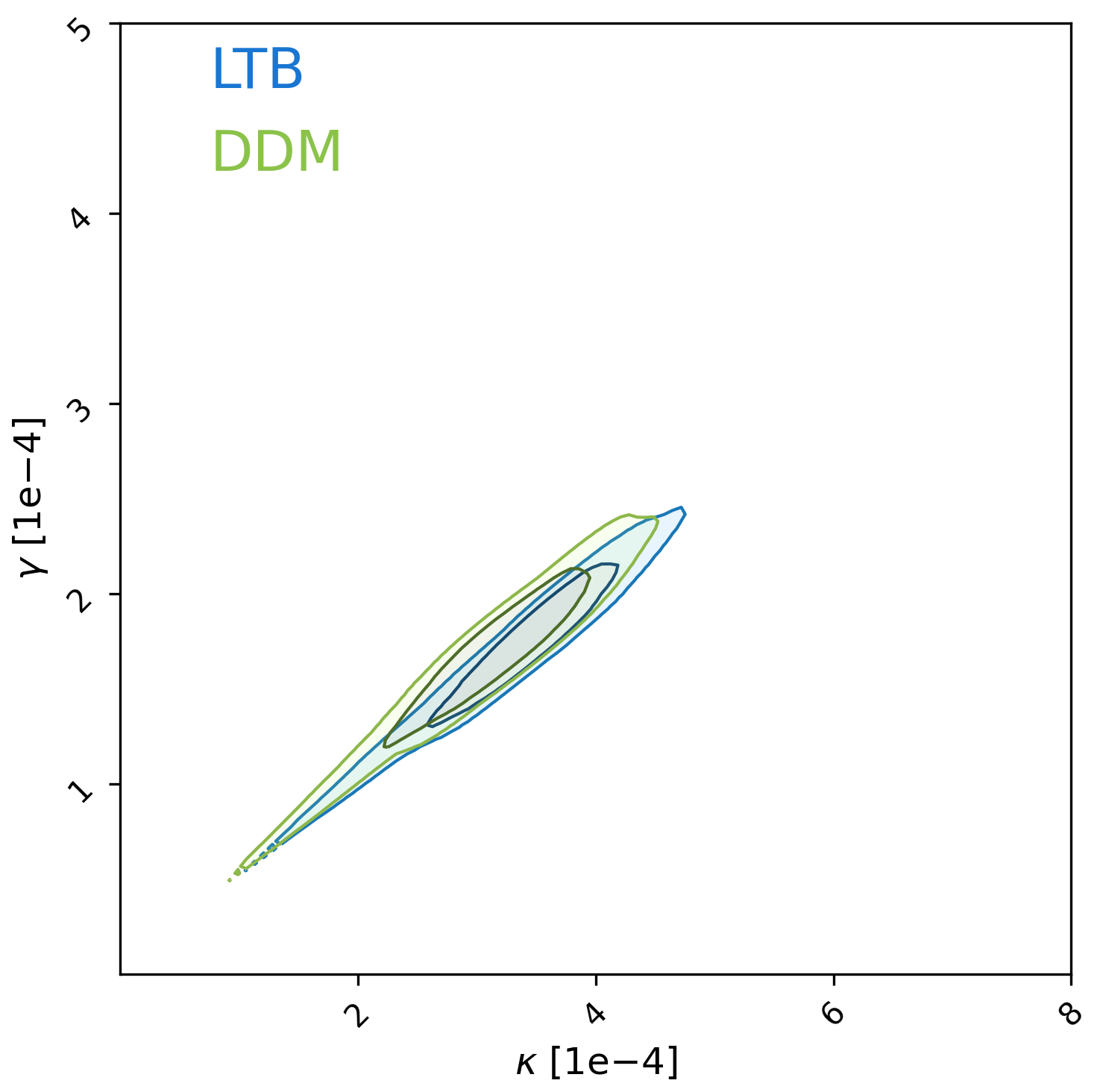}
    \caption{$20 < R_{v} < 30$ Mpc, $v_{in} = 75 \, \text{km}{s}^{-1}$}
    \label{subfig:2.5}
\end{subfigure}
~ 
\begin{subfigure}[t]{0.3\textwidth}
    \centering
    \includegraphics[scale=0.45]{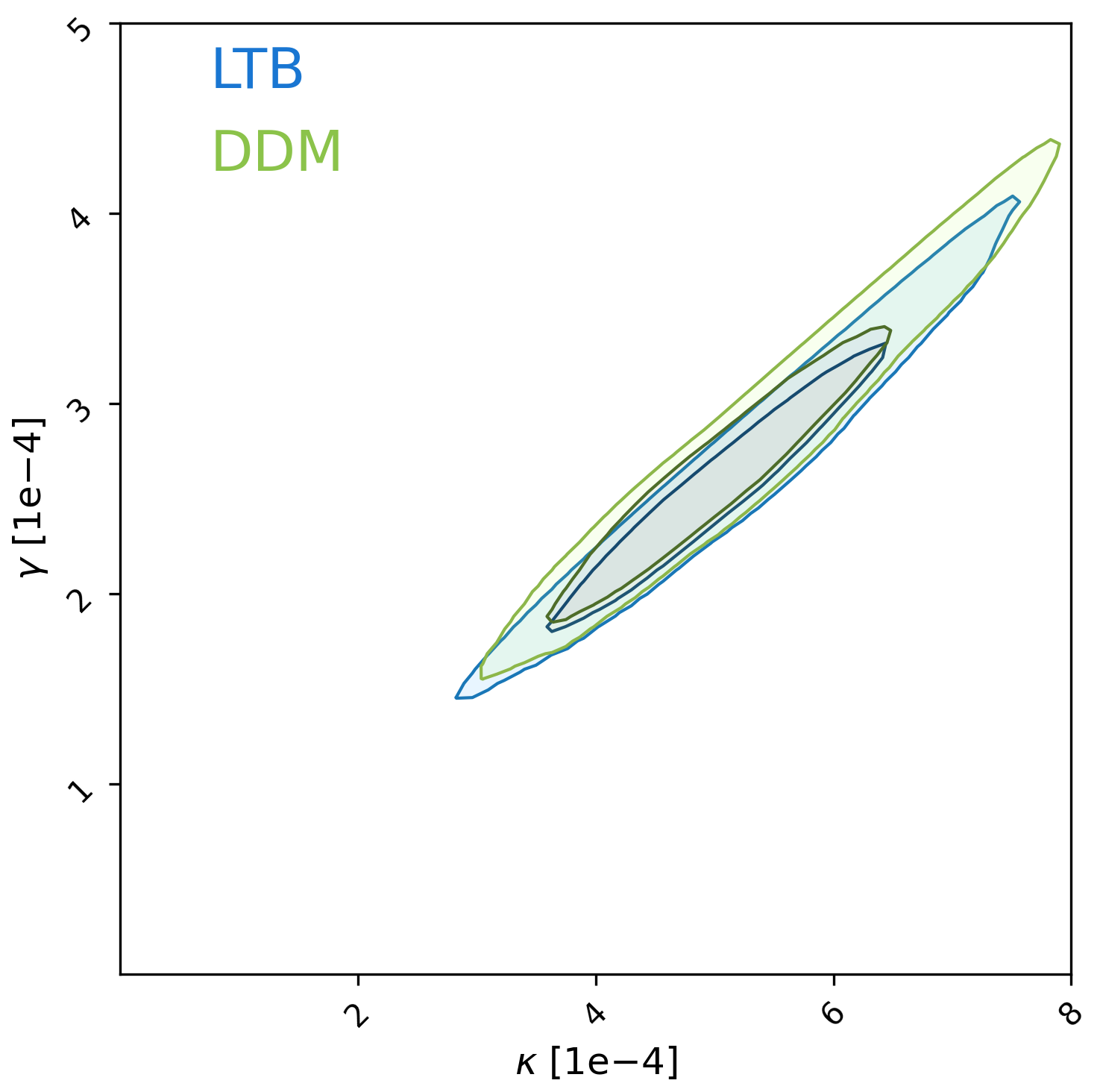}
    \caption{$30 < R_{v} < 40$ Mpc, $v_{in} = 75 \, \text{km}{s}^{-1}$}
    \label{subfig:2.6}
\end{subfigure}
\caption{The distribution of values of maximal amplitude of weak lensing convergence and maximum value of weak lensing shear for voids with DDM and without DDM (i.e. the LTB case), for the three void size classes, the low decay-rate case of $\Gamma = 1.0 \, \text{H}_{0}$ and two injection velocities $v_{in} = 25$ and $75 \, \text{km}\text{s}^{-1}$.}
\label{fig:fig_wl_g1}
\end{figure*}

\begin{figure*}
\centering
\begin{subfigure}[h]{0.3\textwidth}
    \centering
    \includegraphics[scale=0.45]{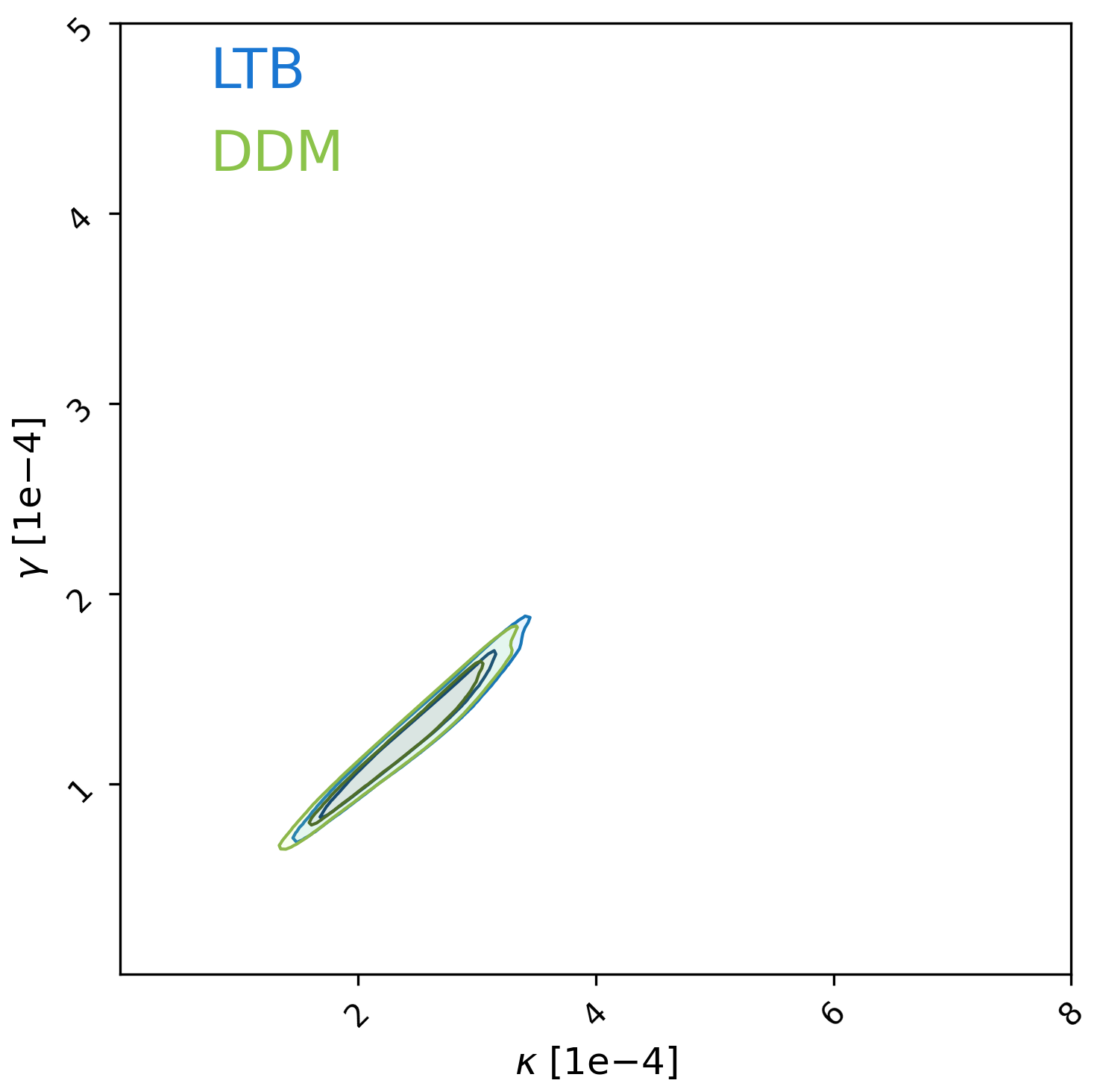}
    \caption{$10 < R_{v} < 20$ Mpc, $v_{in} = 25 \, \text{km}{s}^{-1}$}
    \label{subfig:3.1}
\end{subfigure}%
~ 
\begin{subfigure}[h]{0.3\textwidth}
    \centering
    \includegraphics[scale=0.45]{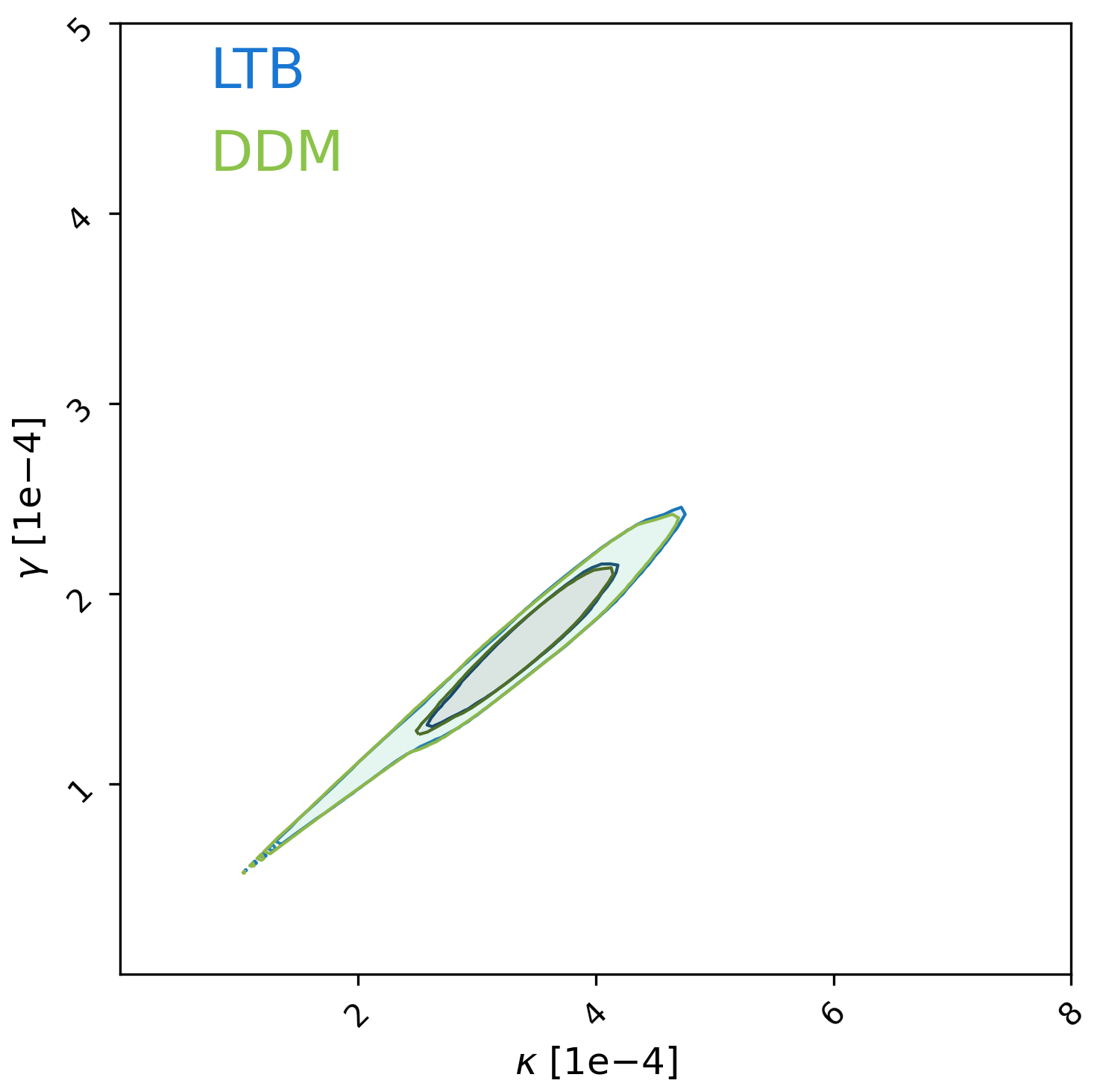}
    \caption{$20 < R_{v} < 30$ Mpc, $v_{in} = 25 \, \text{km}{s}^{-1}$}
    \label{subfig:3.2}
\end{subfigure}
~ 
\begin{subfigure}[h]{0.3\textwidth}
    \centering
    \includegraphics[scale=0.45]{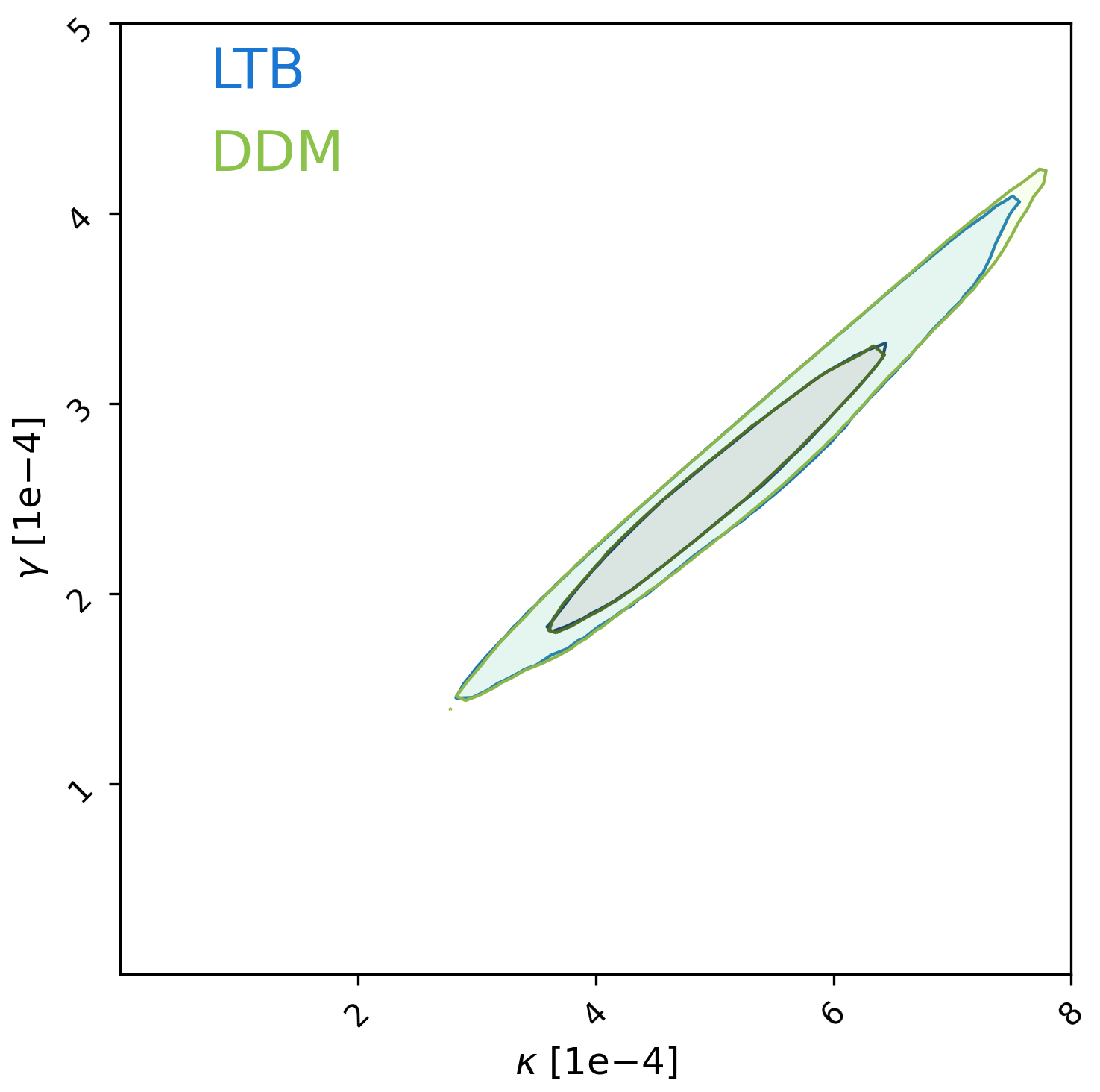}
    \caption{$30 < R_{v} < 40$ Mpc, $v_{in} = 25 \, \text{km}{s}^{-1}$}
    \label{subfig:3.3}
\end{subfigure}
\vfill
\begin{subfigure}[t]{0.3\textwidth}
    \centering
    \includegraphics[scale=0.45]{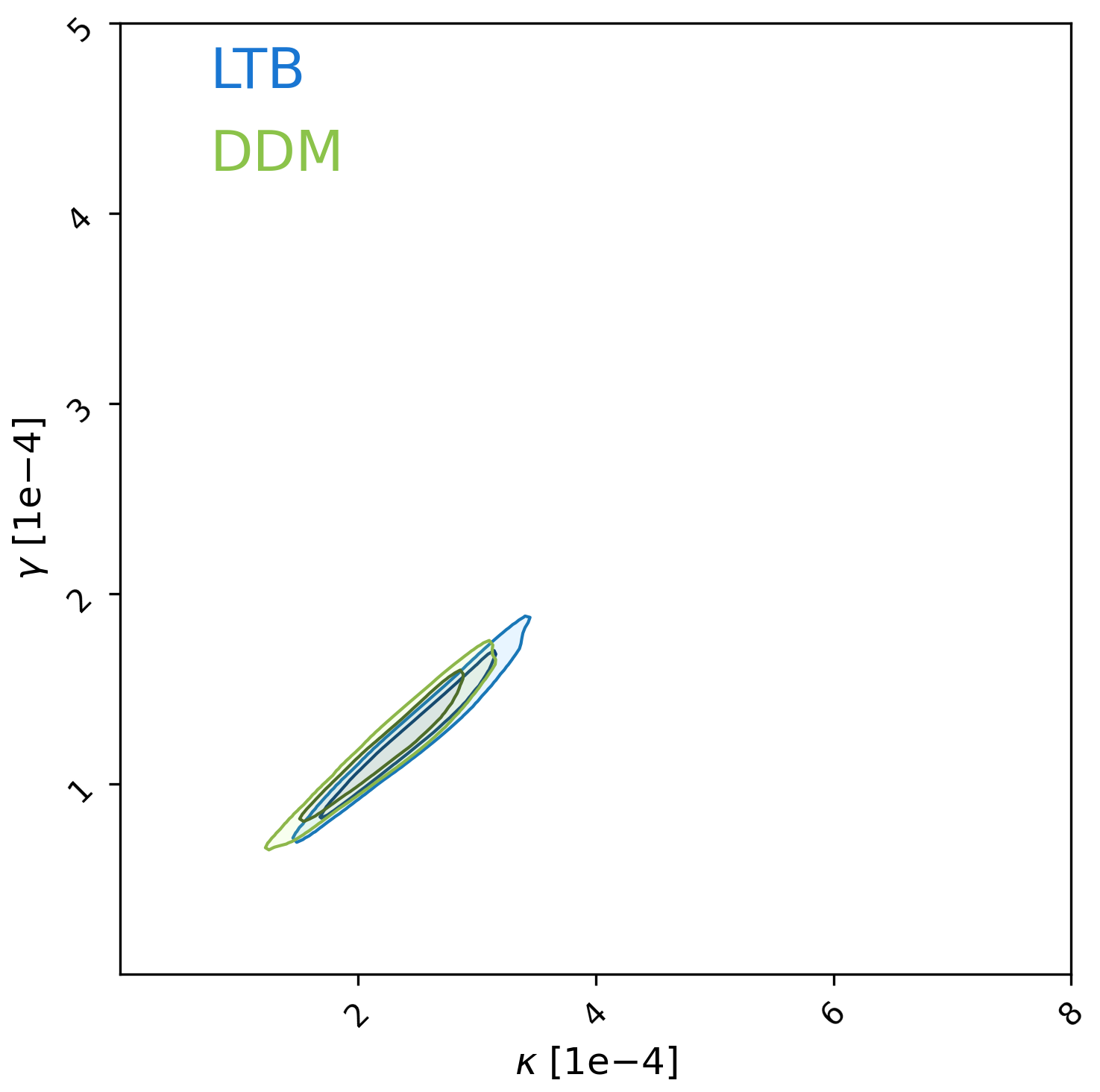}
    \caption{$10 < R_{v} < 20$ Mpc, $v_{in} = 75 \, \text{km}{s}^{-1}$}
    \label{subfig:3.4}
\end{subfigure}%
~ 
\begin{subfigure}[t]{0.3\textwidth}
    \centering
    \includegraphics[scale=0.45]{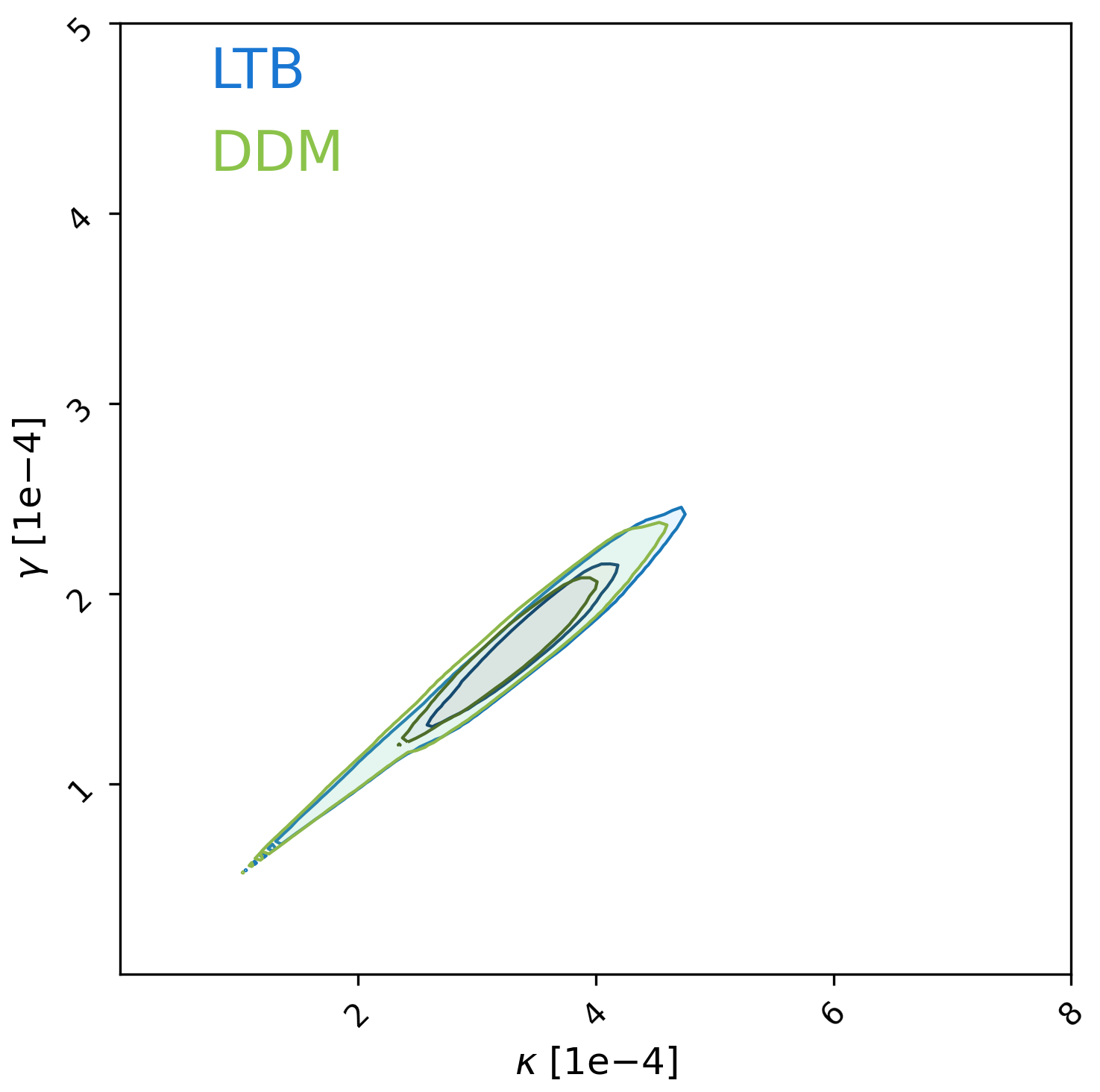}
    \caption{$20 < R_{v} < 30$ Mpc, $v_{in} = 75 \, \text{km}{s}^{-1}$}
    \label{subfig:3.5}
\end{subfigure}
~ 
\begin{subfigure}[t]{0.3\textwidth}
    \centering
    \includegraphics[scale=0.45]{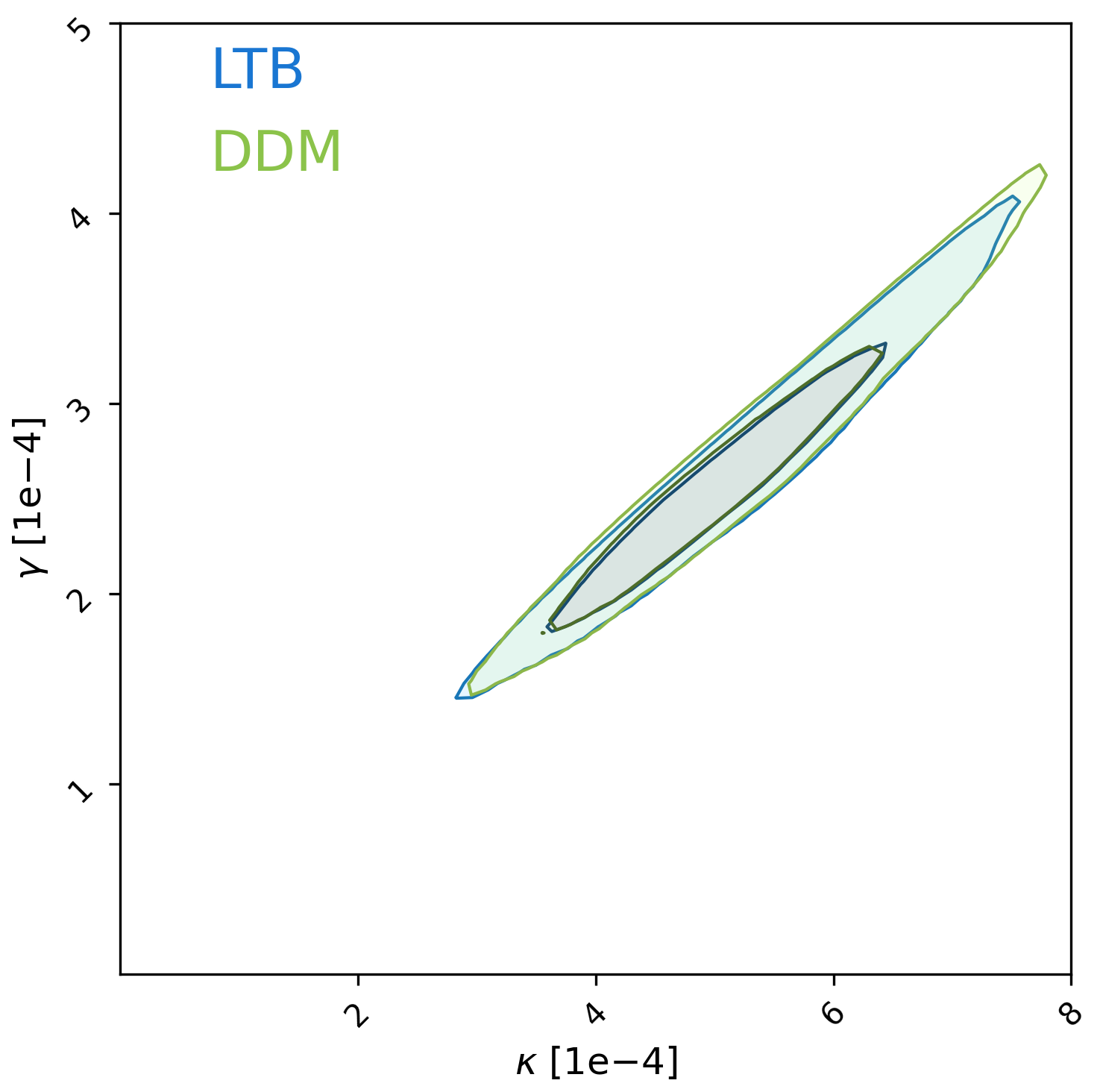}
    \caption{$30 < R_{v} < 40$ Mpc, $v_{in} = 75 \, \text{km}{s}^{-1}$}
    \label{subfig:3.6}
\end{subfigure}
\caption{The distribution of values of maximal amplitude of weak lensing convergence and maximum value of weak lensing shear for voids with DDM and without DDM (i.e. the LTB case), for the three void size classes, the low decay-rate case of $\Gamma = 2.0 \, \text{H}_{0}$ and two injection velocities $v_{in} = 25$ and $75 \, \text{km}\text{s}^{-1}$.}
\label{fig:fig_wl_g2}
\end{figure*}


\begin{figure*}
\centering
\begin{subfigure}[h]{0.3\textwidth}
    \centering
    \includegraphics[scale=0.45]{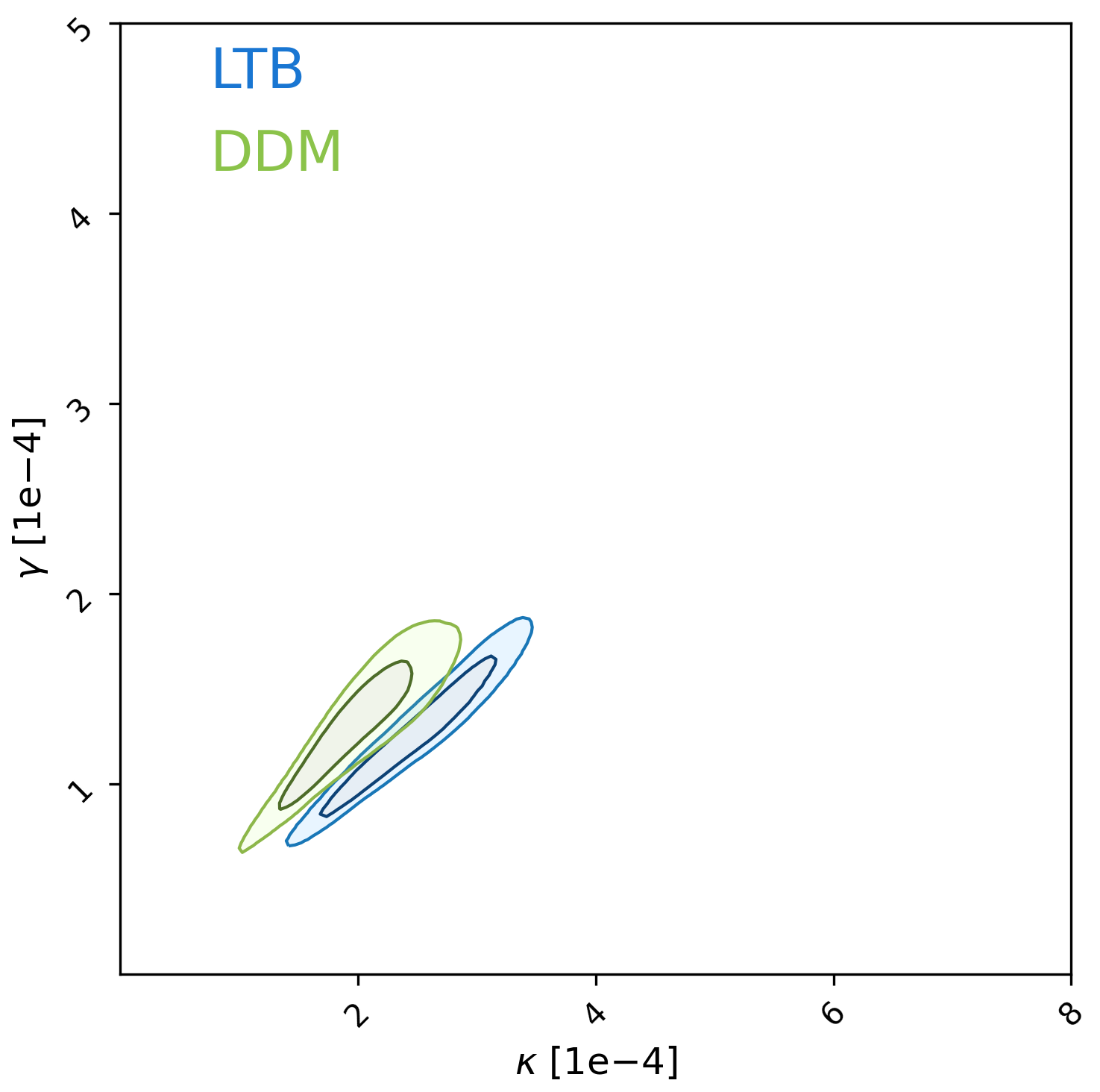}
    \caption{$\Delta \kappa / \kappa \sim 2 \%$}
    \label{subfig:4.1}
\end{subfigure}%
~ 
\begin{subfigure}[h]{0.3\textwidth}
    \centering
    \includegraphics[scale=0.45]{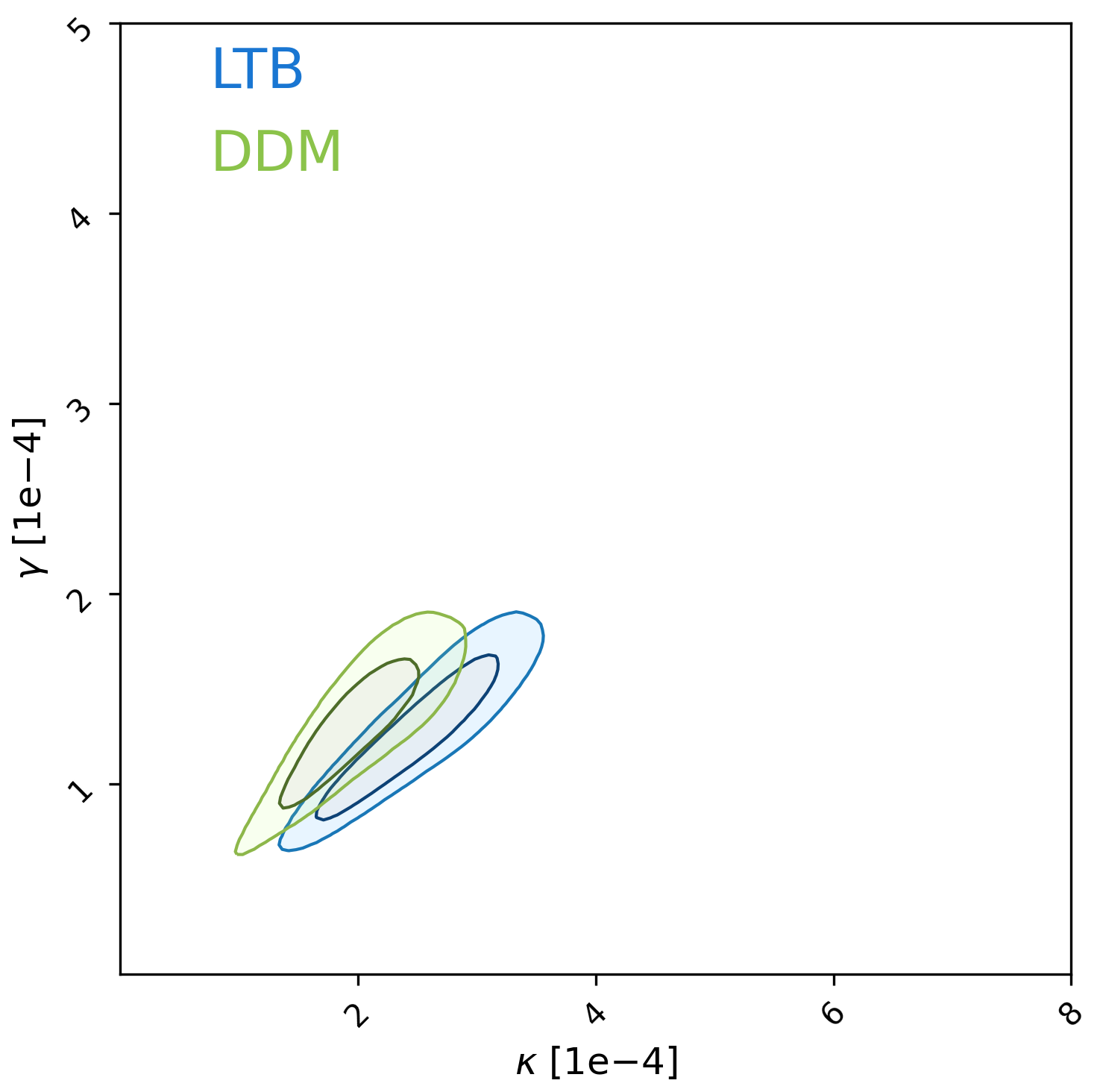}
    \caption{$\Delta \kappa / \kappa \sim 5 \%$}
    \label{subfig:4.2}
\end{subfigure}
~ 
\begin{subfigure}[h]{0.3\textwidth}
    \centering
    \includegraphics[scale=0.45]{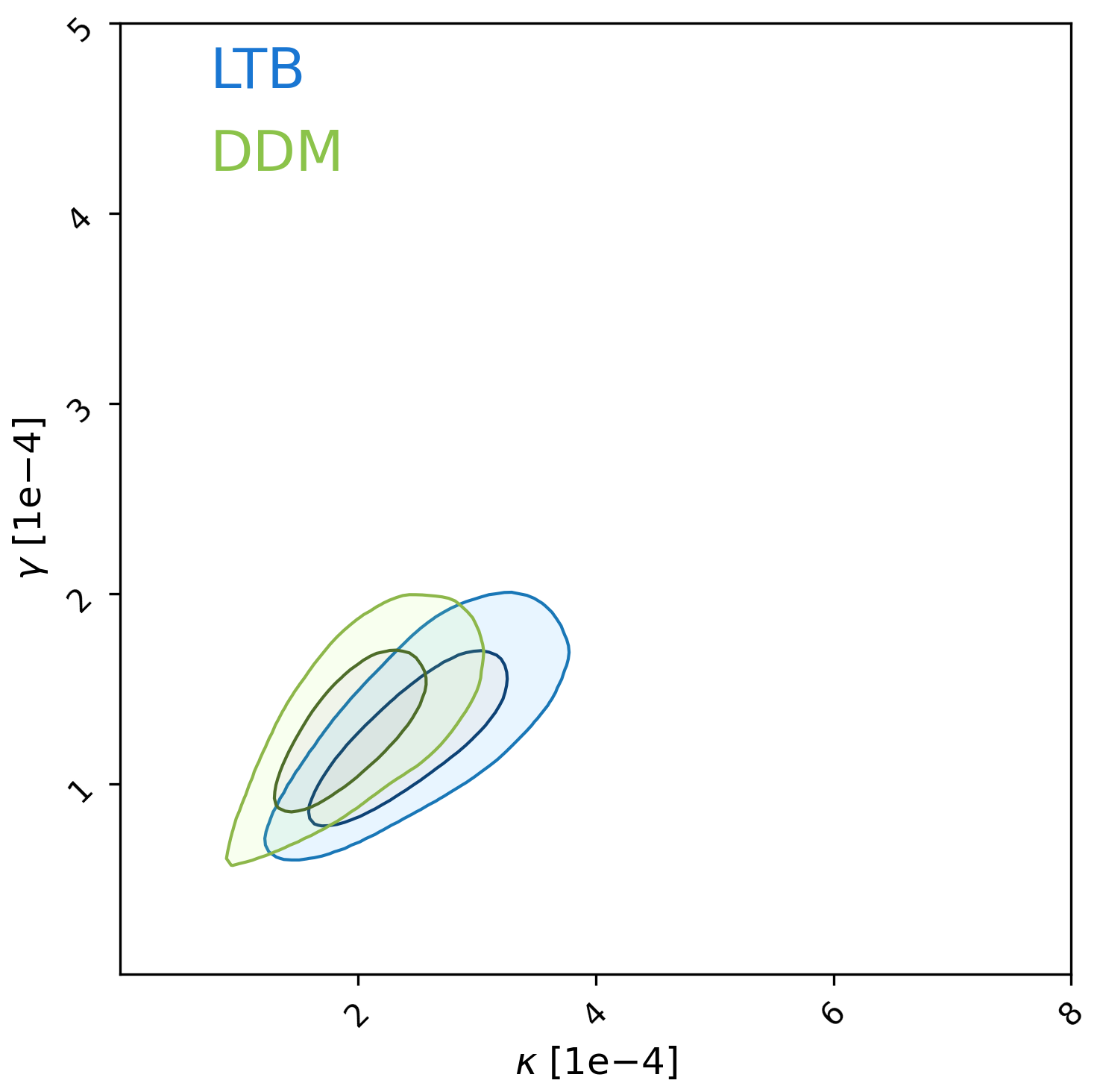}
    \caption{$\Delta \kappa / \kappa \sim 10 \%$}
    \label{subfig:4.3}
\end{subfigure}
\caption{The distribution of values of maximal amplitude of weak lensing convergence and maximum value of weak lensing shear for voids with DDM and without DDM with the introduction of Gaussian noise. The parameters used in this plot are $10 < R_{v} < 20$ Mpc, $v_{in} = 25 \, \text{km}{s}^{-1}$, and $\Gamma = 0.5 \, \text{H}_{0}$.}
\label{fig:fig_wl_w_noise}
\end{figure*}

In this Section we test to see whether weak gravitational lensing could successfully be used to search for DDM signatures inside cosmic voids. As seen from Fig. \ref{fig1}, DDM leads to the formation of shallower voids than in the standard non-decaying, cold dark matter case. 
The effect is sensitive to the mass splitting parameter $\epsilon$: the lighter the mass of the daughter particle the  larger the value of the parameter $\epsilon$ and the larger its velocity. If the mass is comparable ($\epsilon \approx 0$) then the relative velocity of the daughter particle compared to the mother particle is also small, meaning that both fluids have a similar comoving velocity and they both expand away in a similar matter from the centre of the voids towards dense regions surrounding the void.

If the mass of the daughter particle is small with respect to its mother, then this implies that during the decay a larger fraction of the mass of the mother particle is converted to the kinetic energy and subsequently larger velocity. Since there is more mass near the edges of voids than in the centre, this implies that on average more particles will flow into the voids. Consequently, the larger the injection velocity the further into the voids these particles will be able to flood. This leads to smaller amplitude of the density contrast and smaller amplitude of the weak lensing signal. 


These results are summarised in Figs. \ref{fig:fig_wl_g05} - \ref{fig:fig_wl_g2} which show the distribution of the expected measurements of the weak-lensing on cosmic voids based on their size and injection velocity, which is equivalent the mass splitting parameter $\epsilon$. When the velocity of daughter particles is too low to flood the void, there is no discernible difference between voids with and without DDM (top rows in Figs. \ref{fig:fig_wl_g05} - \ref{fig:fig_wl_g2}). 

As the mass split parameter decreases and the relative velocity of daughter particles increase the flooding of voids becomes more efficient. This is especially visible in the case of smaller voids (left hand columns in Figs. \ref{fig:fig_wl_g05} - \ref{fig:fig_wl_g2}).  As the velocity increases the effect become more prominent regardless of the size of the void. Moreover, as the decay rate becomes longer the effect is also lessened. 

Still the effect does not lead to a clear distinction between voids with or without DDM, and it can only be visible with as a statistical effect of deficit of voids with a large amplitude of weak lensing convergence and shear. In addition, the results presented in Fig. \ref{fig:fig_wl_g05} - \ref{fig:fig_wl_g2} assume perfect measurements.

The impact of uncertainties is summarised in Fig. \ref{fig:fig_wl_w_noise} in which increasing degrees of Gaussian noise has been artificially introduced. Figure \ref{fig:fig_wl_w_noise} corresponds to lower left panel of Fig. \ref{fig:fig_wl_g05} where the data were Gaussian scattered with the standard deviation of $2\%$ (left panel), $5\%$ (middle panel), and $10\%$ (right panel). These results show that in order to detect the deficit in the measurements of weak lensing convergence, compared to the expected convergence for a given void with a given shear, we require measurements of weak lensing signal with a precision of $2\%$.  

	
Recent analyses of purely radiative DDM models have shown that contraints of the nature of dark matter can be determined from cosmological data such as the CMB as measured by \textit{Planck}, however, lensing surveys such as \textit{KiDS} were not able to contrain the parameter space significantly, \cite{bucko2022constraining}.

In the work of \cite{wang2012effects} linear modelling showed that unstable dark matter with a minimum injection velocity of $90 \, \text{km}\text{s}^{-1}$ and maximum lifetime of $5 \, \text{Gyr}$ was shown to be possibly observed in forth-coming surveys. As shown in the lower left panels of Fig. \ref{fig:fig_wl_g1} and Fig. \ref{fig:fig_wl_g05}, the non-linear but approximate methods that we have developed has been able to probe a lower limit on the injection velocity down to $75  \, \text{km}\text{s}^{-1}$ and maximum lifetime of the order of the age of the universe, which perhaps with future work, could been shown to be observed in up-coming surveys. 

The analysis here suggests that signatures of DDM with massive daughter particles can be observed in a statistical manner by measuring the gravitational lensing of large number of voids. The precision required to detect DDM signatures is high and beyond the capabilities of current weak lensing surveys but with the forth-coming results of the LSST, \cite{ivezic2019lsst, mahony2022forecasting}, perhaps soon achievable. 



\section{Conclusion} \label{sec:conc}

Over the last 50 years astronomers have acquired overwhelming observational evidence that points to dark matter having a particle nature \cite{2018JCAP...03..026B}, and so represents a lacunae in the standard model of particle physics  \cite{PhysRevLett.126.121802}. To date, there has been no clear and direct detection of dark matter particles, \cite{PhysRevLett.129.161804,PhysRevLett.129.161805} and thus its nature remains unknown. In this paper we have turned to cosmological structure to search for possible signatures of dark matter decay which could then in turn be used to constrain possible properties of dark matter.

This paper is based on the framework developed in \cite{lester2021imprints}, where we showed that the decay of dark matter can lead to distinct features near the edges of cosmic voids. Here we have developed further the model of decay and tested if the DDM signatures could be observed via gravitational lensing of cosmic voids 
\cite{hossen2022mapping,hossen2022ringing}.

Gravitational lensing can probe the overall matter distribution inside cosmic voids. The deeper and larger the void the greater the amplitude of the weak-lensing shear and convergence produced. The decay of dark matter leads to flooding of voids making them more shallow near their edges. This results in both weak-lensing convergence and shear being smaller than in the case of no DDM. The weak-lensing convergence is more sensitive to this flooding than weak lensing shear, and this in the parameter space of weak-lensing convergence and shear, i.e. $\kappa-\gamma$, the DDM shows up as excess of low convergence voids and deficit of voids of high convergence (for a fixed value of shear).

The signatures of DDM can thus be observed in a statistical manner by measuring the gravitational lensing of large number of voids. Still, the precision required to detect DDM signatures is quite high, of the order of few percent, which is beyond the capabilities of current weak-lensing surveys. It is hoped however that including other observables, such as Doppler magnification \cite{bolejko2013antilensing,bacon2014cosmology,hossen2022mapping,hossen2022ringing} could make detection of DDM feasible in the near future.

\section*{Acknowledgements}

The authors would like to thank Andrew Bassom for critically reading preliminary drafts of the manuscript and for offering insightful suggestions.


\bibliographystyle{apsrev}

\end{document}